\documentclass[twocolumn]{aastex63}
\usepackage[utf8]{inputenc}
\usepackage{amsmath}
\usepackage{multirow}
\usepackage{epsfig,lineno}
\usepackage{amsmath,hyperref}
\usepackage{graphicx}
\usepackage{natbib}

\shortauthors{Lewis et. al.}

\begin{document}

\newcommand{\rev}[1]{#1}

\title{Gemini Planet Imager Observations of a Resolved Low-Inclination Debris Disk Around HD 156623}
\correspondingauthor{Briley Lewis}
\email{blewis@astro.ucla.edu}

\author[0000-0002-8984-4319]{Briley L. Lewis}
\affiliation{Department of Physics and Astronomy, UCLA, Los Angeles, CA 90024 USA}

\author[0000-0002-0176-8973]{Michael P. Fitzgerald}
\affiliation{Department of Physics and Astronomy, UCLA, Los Angeles, CA 90024 USA}

\author[0000-0002-0792-3719]{Thomas M. Esposito}
\affiliation{Department of Astronomy, University of California Berkeley, Berkeley, CA 94720 USA}
\affiliation{SETI Institute, Carl Sagan Center, 339 Bernardo Ave, Suite 200, Mountain View, CA 94043, USA}

\author[0000-0001-6364-2834]{Pauline Arriaga}
\affiliation{Arete, 9301 Corbin Ave STE 2000 Northridge, CA 91324}
\affiliation{Department of Physics and Astronomy, UCLA, Los Angeles, CA 90024 USA}

\author[0000-0001-7646-4184]{Ronald L\'{o}pez}
\affiliation{Department of Physics, UCSB, Santa Barbara, CA 93106 USA}
\affiliation{Department of Physics and Astronomy, UCLA, Los Angeles, CA 90024 USA}

\author[0000-0003-4909-256X]{Katie A. Crotts}
\affiliation{Physics \& Astronomy Department, University of Victoria, 3800 Finnerty Rd. Victoria, BC, V8P 5C2, Canada}

\author[0000-0002-5092-6464]{Gaspard Duch\^ene}
\affiliation{Univ. Grenoble Alpes, CNRS, IPAG, F-38000 Grenoble, France}
\affiliation{Department of Astronomy, University of California Berkeley, Berkeley, CA 94720 USA}

\author[0000-0002-7821-0695]{Katherine B. Follette}
\affiliation{Department of Physics and Astronomy, Amherst College, Amherst, MA 01002 USA}

\author[0000-0001-9994-2142]{Justin Hom}
\affiliation{Steward Observatory and Department of Astronomy, University of Arizona, Tucson AZ 85721 USA}

\author[0000-0002-6221-5360]{Paul Kalas}
\affiliation{Astronomy Department, University of California, Berkeley, CA 94720, USA}
\affiliation{SETI Institute, Carl Sagan Center, 339 Bernardo Ave, Suite 200, Mountain View, CA 94043, USA}
\affiliation{Institute of Astrophysics, FORTH, GR-71110 Heraklion, Greece}

\author[0000-0003-3017-9577]{Brenda C. Matthews}
\affiliation{Herzberg Astronomy and Astrophysics, National Research Council of Canada, 5071 West Saanich Road, Victoria, BC V9E 2E7, Canada}
\affiliation{Physics \& Astronomy Department, University of Victoria, 3800 Finnerty Rd. Victoria, BC, V8P 5C2, Canada}

\author[0000-0002-8382-0447]{Maxwell Millar-Blanchaer}
\affiliation{Department of Physics, University of California Santa Barbara, Santa Barbara, CA 93106 USA}

\author[0000-0003-1526-7587]{David J. Wilner}
\affiliation{Center for Astrophysics, Harvard \& Smithsonian, 60 Garden Street, Cambridge, MA 02138-1516, USA}

\author[0000-0002-9133-3091]{Johan Mazoyer}
\affiliation{LESIA, Observatoire de Paris, Université PSL, Sorbonne Université, Université Paris Cité, CNRS, 5 place Jules Janssen, 92195 Meudon, France}

\author{Bruce Macintosh}
\affiliation{University of California Santa Cruz, 1156 High St, Santa Cruz, CA 95064 USA}
\affiliation{Kavli Institute for Particle Astrophysics and Cosmology, Department of Physics, Stanford University, Stanford, CA 94305 USA}





\begin{abstract}

The 16 Myr-old A0V star HD 156623 in the Scorpius--Centaurus association hosts a high-fractional-luminosity debris disk, recently resolved in scattered light for the first time by the Gemini Planet Imager (GPI) \rev{in polarized intensity}. We present new analysis of the GPI \textit{H}-band polarimetric detection of the HD 156623 debris disk, with particular interest in its unique morphology. This debris disk lacks a visible inner clearing, unlike the majority of low-inclination disks in the GPI sample and in Sco-Cen, and it is known to contain CO gas, positioning it as a candidate ``hybrid'' or ``shielded'' disk. We use radiative transfer models to constrain the geometric parameters of the disk based on scattered light data and thermal models to constrain the unresolved inner radius based on the system's spectral energy distribution (SED). We also compute a measurement of the polarized scattering phase function, adding to the existing sample of empirical phase function measurements. We find that HD 156623's debris disk inner radius \rev{is constrained to less than 26.6 AU from scattered light imagery and less than 13.4 AU from SED modeling at a 99.7\% confidence interval}, and suggest that gas drag may play a role in retaining sub-blowout size dust grains so close to the star. 

\end{abstract}

\keywords{planets and satellites: formation; protoplanetary disks; infrared: planetary systems; polarization; high contrast imaging; methods: data analysis; debris disks}

\section{Introduction} \label{sec:intro}

High-contrast imaging is extremely useful for spatially resolving debris disks, the extrasolar analogs of our solar system's interplanetary dust/Kuiper Belt \citep{mann2006dust}. Debris disks represent a later stage of \rev{a system's formation and evolution}, wherein the circumstellar disk has evolved and lost most of its gas content. They also have low optical depth, and their dust is thought to be replenished by collisions and/or sublimation of planetesimals \citep{wyatt2008evolution}. \rev{Imaging of debris disks provides significant information about the system beyond what an infrared excess measurement alone can provide, such as grain properties and disk morphology \citep{hughes2018debris}. Longer-wavelength observations (e.g. those from ALMA, the Atacama Large Millimeter Array) are able to resolve the thermal emission of disks, providing information on the distribution of large grains and the presence of gas in a system. In the near-infrared and optical, however, we are able to trace smaller grains, and polarimetric scattered light observations can reveal additional information about a disk's composition and structure.} Morphological information can even reveal signs of a planetary companion, as demonstrated by the disk warp that led to the discovery of $\beta$ Pictoris b \citep{roques1994there,2010Sci...329...57L}.

The Gemini Planet Imager (GPI) \citep{macintosh2008gemini, Macintosh12661}, a ground-based adaptive-optics (AO) instrument dedicated to high-contrast discovery and characterization of exoplanets (formerly at Gemini South), has also imaged multiple debris disks with its integral-field spectrometer and dual-channel polarimeter \citep{esposito2020debris}. In addition to the planet search component of the Gemini Planet Imager Exoplanet Survey (GPIES), whose results are reported in \citet{nielsen2019gemini}, \citet{esposito2020debris} reports on a survey of debris disks with GPI. This survey resolved 26 debris disks and 3 protoplanetary/transitional disks, several of which were resolved in scattered light for the first time. 

One of these newly resolved disks was HD 156623, a debris disk around a 16 Myr-old A0V star in Sco-Cen, at a distance of $\sim$112 pc \citep{houk1982vol, 2018yCat.1345....0G}. HD 156623 is a low-inclination disk \rev{($\approx$ 35$^\circ$)}, detected by GPI only in polarized light \citep{esposito2020debris}. This system lacks a visible inner clearing unlike many other imaged debris disks, including others in Sco-Cen \citep{bonnefoy2021narrow}; close to the star, dust is usually cleared by Poynting-Robertson drag, accretion, stellar radiation pressure, and/or blowout from stellar winds on short timescales \citep{wyatt2008evolution,hughes2018debris}, and a planet, if present, can sculpt a sharp inner edge of a debris disk (e.g. \citet{engler2020hd,pearce2024effect}). Additionally, prior observations in \rev{thermal emission via ALMA imaging from} \citet{lieman2016debris} reported evidence for an inner clearing and showed significant CO gas mass, leading to the ``hybrid'' disk label given to this system. There is still significant debate on the mechanism that leads to the observed gas mass; either primordial gas has been retained (i.e. due to shielding by carbon atoms from photodissociation of CO) \citep{kral2019imaging,moor2019new,marino2020population} or secondary gas has been re-generated via collisions, sublimation, or another process \citep{pericaud2017hybrid, hales2019modeling, moor2017molecular,beust1990beta}.


In this work, we aim to perform more detailed modeling of the disk around HD 156623 as seen in polarized infrared scattered light by GPI, with particular interest in constraining the presence and location of the hidden inner edge of the debris disk. In Sections \ref{subsec:hd156623} and \ref{sec:obs} respectively, we include information on the host star HD 156623 and observations done by the Gemini Planet Imager with relevant information on the data reduction. In Section \ref{subsec:MCFOST}, we present \rev{scattered light imagery} models from \texttt{MCFOST} and the resulting best-fit parameters for the disk geometry \citep{pinte2006monte,pinte2009benchmark,pymcfost}. In Section \ref{sec:bright}, we present empirical measurements of radial brightness profiles, scattering phase function, and polarization fraction. We then present analysis of the system's spectral energy distribution (SED) in Section \ref{sec:SED}. These measurements \rev{and insights into the  inner edge location and minimum grain size in the context of this gas-rich disk} are then discussed in Section \ref{sec:disc}.


\section{Previous Observations of HD 156623} \label{subsec:hd156623}

HD 156623 (HIP 84881) is an A0V star with a parallax of 9.2$\pm$0.03 mas \citep{houk1982vol, 2018yCat.1345....0G, brown2021gaia}. Stellar parameters are summarized in Table \ref{tab:star}. It exhibits pulsations consistent with zero-age main sequence (ZAMS) $\delta$ Scutis stars, and it is a member of the Upper Centaurus-Lupus subgroup of the Sco-Cen association, which contains stars of a wide-range of ages from $\sim$3 to 19 Myr \citep{pecaut2016star,mellon2019discovery,ratzenbock2023star}. \rev{This star system appears to be relatively young} --- 16$\pm$7 Myr according to \citet{mellon2019discovery}. 

\begin{table}
\begin{centering}
\begin{tabular}{|l|c|l|}
\hline
\textbf{Parameter} & \textbf{Value} & \textbf{Reference} \\ \hline
D (pc) & 111.75 $\pm$ 0.96 & \citet{gaia2018gaia} \\ \hline
Age (Myr) & 16 $\pm$ 7 & \citet{mellon2019discovery} \\ \hline
T$_\text{eff}$ (K) & 9040$^{+240}_{-160}$ & \citet{mellon2019discovery} \\ \hline
M$_*$ (M$_\odot$) & 1.90$^{+0.05}_{-0.04}$ & \citet{esposito2020debris} \\ \hline
L$_*$ (L$_\odot$) & 13.06 $\pm$ 1.80 & \citet{esposito2020debris} \\ \hline
R$_*$ (R$_\odot$) & 1.51 $\pm$ 0.09 & \citet{mellon2019discovery} \\ \hline
$I$ (mag) & 7.1 & \citet{esposito2020debris} \\ \hline
$H$ (mag) & 7.0 & \citet{esposito2020debris} \\ \hline
$L_\text{IR}$/$L_*$ & 4.3$\cdot 10^{-3}$ & \citet{rizzuto2012WISE} \\ \hline
\end{tabular}
\caption{Summary of parameters for HD 156623. Magnitudes \rev{are synthetic photometry in Cousins $I$ \citep{1998A&A...333..231B} and 2MASS $H$ \citep{2006AJ....131.1163S} as calculated by \citet{esposito2020debris}.}}\label{tab:star}
\end{centering}
\end{table}


HD 156623 has been covered by many ground-based photometric surveys and space-based missions--including \textit{WISE} \citep{cutri2012vizier,wright2010wide}, 2MASS \citep{2006AJ....131.1163S}, \textit{IRAS} \citep{1984ApJ...278L...1N, beichman1988infrared}, and Tycho \citep{2000A&A...355L..27H,2010PASP..122.1437P}--which is useful for constructing an SED as described in Section \ref{sec:SED}. Data from \textit{WISE} indicate an infrared excess for HD 156623 of $L/L_* =$ 4.33$\times 10^{-3}$ \citep{rizzuto2012WISE}. 

The disk around HD 156623 was also previously observed with ALMA, which marginally resolved the disk and found a strong $^{12}$CO(2-1) signature, indicating the presence of a mass of CO gas $\geq 3.9 \cdot 10^{-4} M_\oplus$ \citep{lieman2016debris}. \citet{lieman2016debris} also reports a 1240 $\mu$m continuum flux for HD 156623 of 720$\pm$110 $\mu$Jy. Further detections of CO gas with APEX and IRAM were also completed in \citet{pericaud2017hybrid}. 

\section{Observations and Data Reduction} \label{sec:obs}

\begin{table}
\centering
\begin{tabular}{|l|l|l|r|l|l|}
\hline
\textbf{Date (UT)} & \textbf{Mode} & \textbf{$t_\text{exp}$ (s)} & \textbf{$t_\text{int}$ (s)} & $\Delta$PA ($^\circ$)\\\hline
2019 Apr 27                 & Pol         & 88.74                  & 2129.81      &     28.2     \\\hline
2017 Apr 21                & Pol        & 59.65                  & 954.34           &   11.2   \\\hline
\end{tabular}
\caption{GPI observations of HD 156623. The debris disk was detected in both polarimetric observations, and for the purposes of this investigation we focus on the higher-quality, longer-exposure data from 2019. The following observation parameters are provided: exposure time ($t_\text{exp}$), total integration time ($t_\text{int}$), and field rotation ($\Delta$PA,  degrees). All observations were taken in the \textit{H} band.}\label{tab:obs}
\end{table}

\subsection{Gemini Planet Imager Observations} \label{subsec:GPI}

GPI observations resolved the disk in scattered light using GPI's polarimetric mode, which uses a half-wave plate for modulation and a Wollaston prism beam splitter for analysis \citep{perrin2010imaging,perrin2014gemini}. \rev{The polarimetric data presented herein were first published as part of a larger survey in} \citet{esposito2020debris}. In total intensity, GPI uses angular differential imaging (ADI) to create angular diversity that can be exploited to improve the images' final contrast \citep{marois2006angular}. Observations in pol mode result in measurements of the Stokes $\mathcal{I}$, $\mathcal{Q}$, and $\mathcal{U}$ vectors. Most data products for debris disks --- including those used in this work --- use azimuthal Stokes vectors, $\mathcal{Q}_\phi$ and $\mathcal{U}_\phi$ (described below in Equations \ref{radialstokes} and \ref{radialstokes2}, where $\phi$ is the azimuthal angle around the star measured counterclockwise starting from North), which is useful for understanding polarization relative to the central star \citep{schmid2006limb,millar2015beta}. 

\begin{eqnarray}\label{radialstokes}
    \mathcal{Q}_\phi = \mathcal{Q}\text{cos}2\phi + \mathcal{U}\text{sin}2\phi \\
    \mathcal{U}_\phi = -\mathcal{Q}\text{sin}2\phi + \mathcal{U}\text{cos}2\phi \label{radialstokes2}
\end{eqnarray}

$\mathcal{Q}_\phi$ then contains all polarized light from the debris disk that is aligned perpendicular or parallel to a vector from the host star to a given pixel. $\mathcal{U}_\phi$ then contains polarized light oriented $45^\circ$ to this vector, which we expect to be devoid of signal for an optically thin debris disk (discussed further in \citet{esposito2020debris}). GPI images also include four ``satellite'' spots, reference images of the occulted star used for astrometric and photometric calibrations \citep{sivaramakrishnan2006astrometry,esposito2020debris}.

The effective outer radius of the full circular field of view in reduced images is 1.4$''$ from the star, with radial separations up to 1.8$''$ visible over limited ranges of position angle (PA). The field of view size of each frame is 2.6$''$ x 2.6$''$, and we use the GPI pixel scale value of 14.166 $\pm$ 0.007 mas lenslet$^{-1}$ \citep{konopacky2014gemini,de2015astrometric,de2020revised}. The minimum projected separation viewable in the data is 0.123$''$, set by the edge of the focal plane mask (FPM, \citet{de2015astrometric}).

We observed HD 156623 on 2017 Apr 21 and 2019 Apr 27. A summary of the observations are listed in Table \ref{tab:obs}. The debris disk around HD 156623 was detected by GPI in \textit{H} band ($\lambda_c$=1.65 $\mu$m) polarized light ($\mathcal{Q}_\phi$), in both 2017 and 2019. In 2017, sixteen 60-second exposures were taken in \textit{H} band polarimetry mode with the wave plate rotating between exposures (typically 22.5$^\circ$ per move as described by \citet{perrin2015polarimetry}), with 11.2$^\circ$ of field rotation and approximately 1.15 arcsec seeing. In 2019, twenty four 88-second exposures were taken in \textit{H} band polarimetry mode with 28.2$^\circ$ of field rotation and $\sim$0.39 arcsec seeing. The 2019 data was higher quality and is what will be used in our analysis.  

\subsection{Data Reduction and Polarized Differential Imaging}

Data for HD 156623 were reduced using the standard GPI Data Reduction Pipeline (DRP), documented in \citet{perrin2014gemini}, \citet{perrin2016gemini}, and \citet{wang2018automated}. The data reduction procedure for polarimetry data in particular, as well as sources of noise in GPI polarimetric data, are further described in \citet{perrin2015polarimetry} and Section 4.1 of \citet{esposito2020debris}; this reduction, however, is an improved reduction from that presented in \citet{esposito2020debris} due to the quadropole correction described later in this section. The GPI DRP completes dark subtraction, flexure correction, field distortion correction, destriping, bad pixel corrections, and other calibrations specific to the GPI instrument, such as a cleaning procedure to account for biasing between orthogonal polarization channels \citep{perrin2014gemini,perrin2015polarimetry,wiktorowicz2014gemini, konopacky2014gemini}. After the standard DRP, the series of reduced polarimetric observations is converted into a single data cube containing the Stokes vectors $\mathcal{I}$, $\mathcal{Q}$, $\mathcal{U}$, and $\mathcal{V}$, before then converting $\mathcal{Q}$ and $\mathcal{U}$ to radial Stokes vectors ($\mathcal{Q}_\phi$ and $\mathcal{U}_\phi$) as previously described \citep{perrin2015polarimetry}. This data reduction pipeline produces images in units of analog-to-digital units per coadd (ADU coadd$^{-1}$); using the measured flux of the satellite spots, which are themselves images of the host star, and the star's known \textit{H} band magnitude (see Table \ref{tab:star}) we converted those detector units into surface brightness units of mJy arcsec$^{-2}$ for the subsequent analysis, as in \citet{wang2014gemini}, \citet{hung2016gemini}, and \citet{esposito2020debris}. The calibration factor for this data reduction was \rev{6.41$\times 10^{-9} \pm$ 0.27$\times 10^{-9}$ Jy/(ADU/coadd).} Additionally, an extra instrumental polarization subtraction--a quadrupole modulation of the total intensity pattern scaled to data and subtracted for each waveplate modulation--was applied to the data to remove an instrumental polarization effect seen in GPI images \citep{2020SPIE11447E..5AD}. 


\begin{figure}
    \centering
    \includegraphics[width=\linewidth]{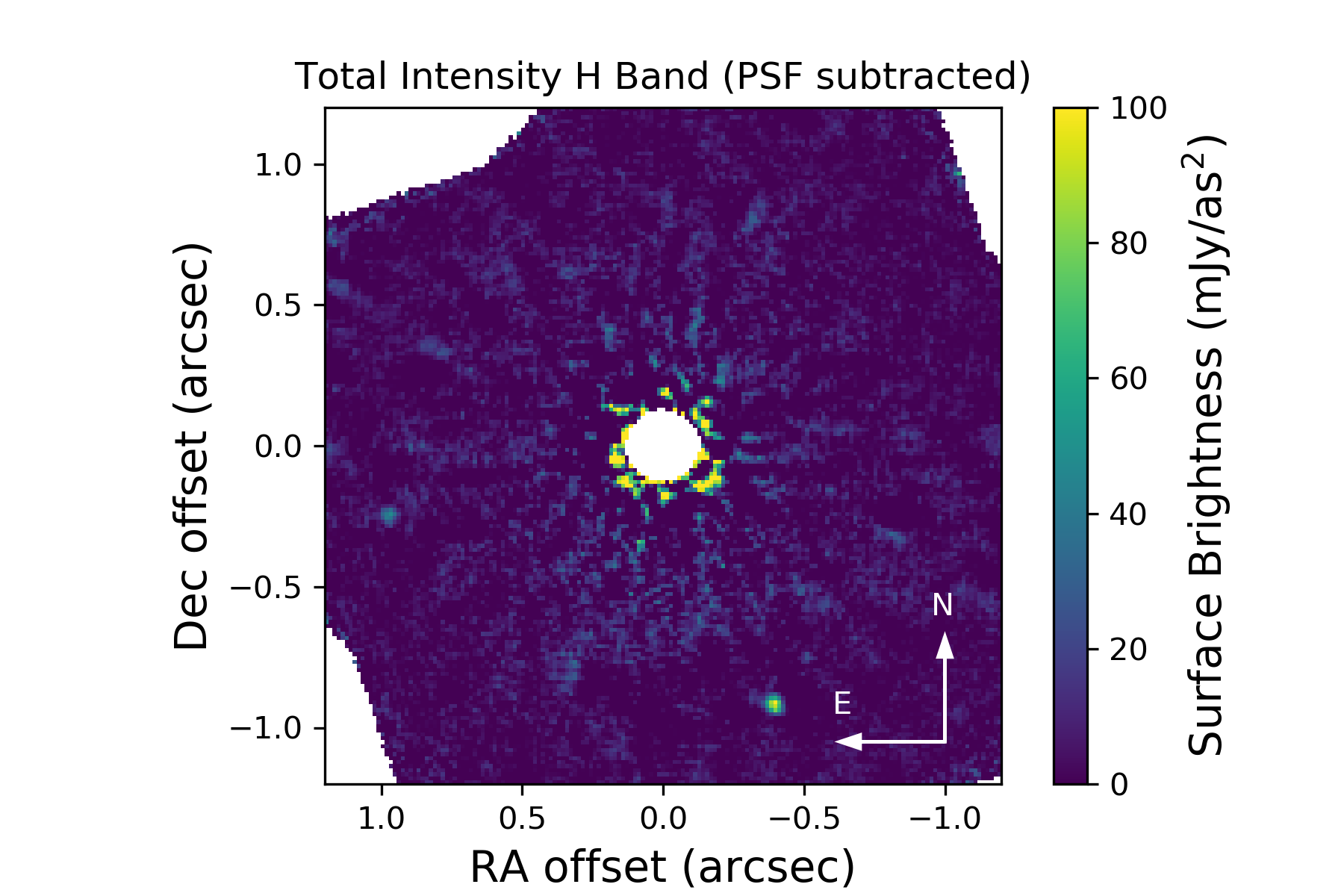}
    \includegraphics[width=\linewidth]{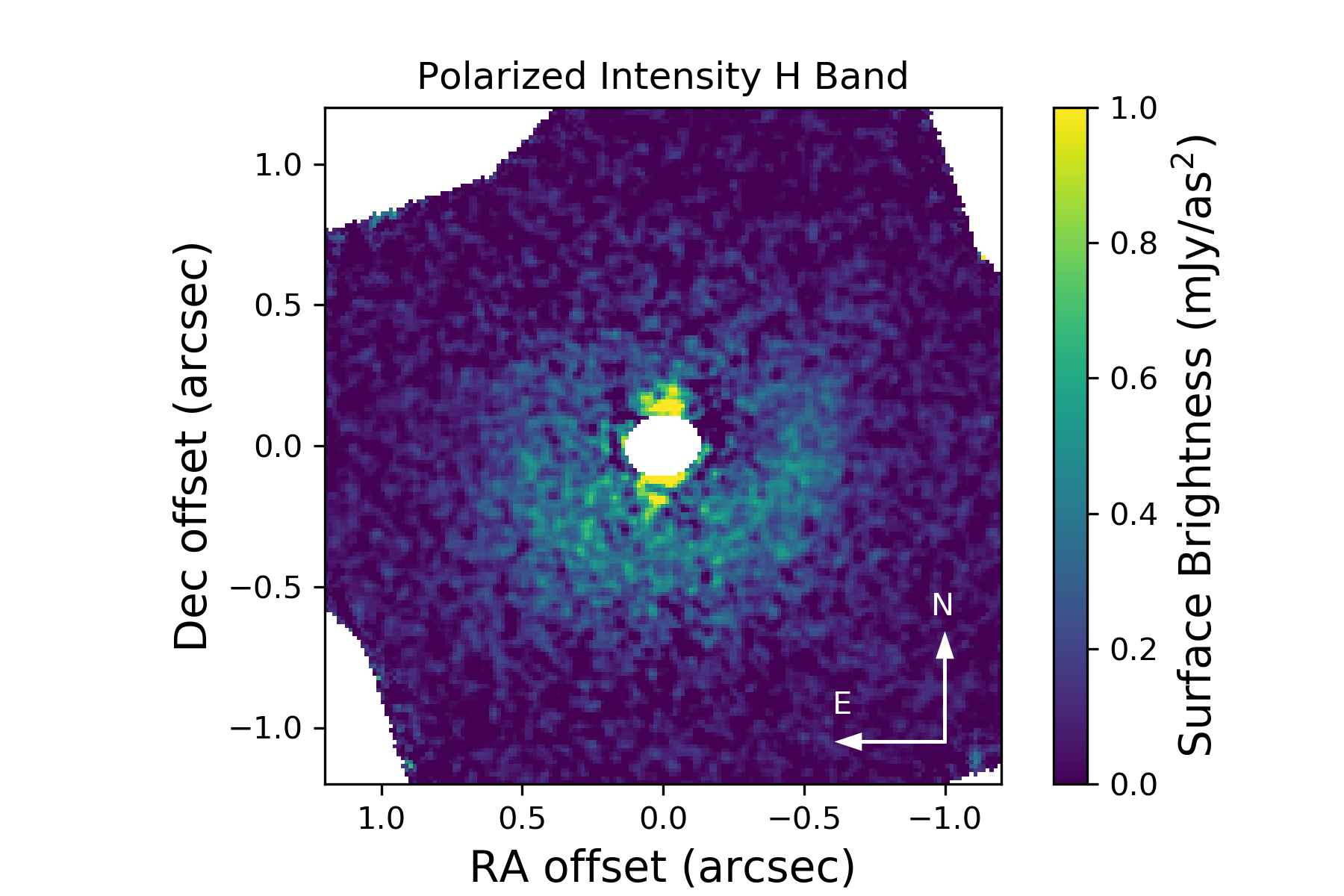}
    \caption{PSF-subtracted total intensity GPI \textit{H} band image of HD 156623 using 5 KL modes (top) and polarized intensity  GPI \textit{H} band image of HD 156623 smoothed with a 2 pixel wide Gaussian kernel for visualization (bottom). The debris disk is only detected in polarized light, with a signal-to-noise ratio of $\sim$2-3 per resolution element. The point sources visible in the total intensity image were determined to be unbound background stars in \citet{esposito2020debris}.}
    \label{fig:raw}
\end{figure}

Polarized intensity frames from the standard data reduction readily show the disk, due to the natural separation of unpolarized stellar light and polarized light scattered from the disk. For total intensity, we must complete stellar point spread function (PSF) subtraction, for which we use the \texttt{pyKLIP} package \citep{wang2015pyklip}. \texttt{pyKLIP} uses Karhunen-Lo\'eve Image Processing (KLIP), a form of principal components analysis, to create a model of the stellar PSF using its dominant modes of variation for PSF subtraction \citep{soummer2012KLIP}. For polarimetric GPI data, the total intensity is the sum of the two orthogonal polarization states, plus any unpolarized light. We employ ADI in our reduction for total intensity data taken with GPI polarimetry mode. Parameters for the \texttt{pyKLIP} reduction generally must be chosen with care to maximize the signal from the disk, avoiding self-subtraction \citep{milli2012impact} while also reducing contamination from the stellar PSF. For this particular disk, which is faint relative to the stellar halo in total intensity, we use a particularly unaggressive implementation of KLIP, using one annulus and one subsection in the subtraction as recommended by \texttt{pyKLIP} documentation. We additionally test multiple KL modes (1, 5, 10, 20, 50) and use \texttt{pyKLIP}'s minimum rotation (\texttt{minrot} = 3) criterion for basis vector selection, as is recommended to avoid self-subtraction. Even after applying PSF-subtraction, however, we were unable to recover the disk in total intensity light, as shown in Figure \ref{fig:raw}. \rev{Although other recent algorithms have focused on improving this issue of self-subtraction (e.g. non-negative matrix factorization (NMF), \citet{ren2018non}), we do not pursue additional reductions in this work; although such algorithms may be able to recover the disk's general morphology, we choose to focus on the polarized intensity detection which already provides such morphological information. Additionally, low-inclination and radially extended disks are particularly challenging to retrieve in total intensity with an ADI approach.} The point sources in total intensity were already determined by \citet{esposito2020debris} to be unbound background objects and are thus not considered in our analysis of this system.

\section{Disk Morphology}\label{subsec:MCFOST}

In this section, we reproduce the modeling originally presented in \citet{esposito2020debris} to show the scattered light model disk images and quality of fit to the data, which were not provided in that work, \rev{with the goal of constraining the inner edge from our imagery.} This modeling is nearly identical to that of \citet{esposito2020debris}, but run with a 40\% longer chain. This leads to a small improvement in the quality of the results due to more total samples of the posterior. We use these morphological results (namely position angle, inclination, and radius constraints) to guide our measurements of the disk's brightness profiles and phase function, calculated in Section \ref{sec:bright}.




As in \citet{esposito2020debris}, morphological parameters of the disk were estimated by comparison to models from the Monte Carlo radiative transfer code \texttt{MCFOST} \citep{pinte2006monte,pinte2009benchmark}. \texttt{MCFOST} is a radiative transfer code that uses Monte Carlo and ray tracing methods to produce models of observables (scattered-light imagery, millimeter and infrared visibilities, line maps, SEDs, dust temperature distributions) for circumstellar environments such as protoplanetary and debris disks. Models were fit to Stokes $\mathcal{Q}_\phi$, $\mathcal{Q}$ and $\mathcal{U}$ data simultaneously, as $\mathcal{Q}_\phi$ is the only available data with the quadrupole noise signal removed. Following the procedure in \citet{esposito2018}, our models assume Mie scattering from spherical grains  in an azimuthally-symmetric, optically-thin disk centered on the star, observed at 1.65 microns (the \textit{H} band peak wavelength). We additionally use the same disk parameters and procedure as described in \citet{esposito2020debris}\rev{, as summarized below. The disk has a vertical structure where the scale height is a constant fraction of the radius, set to 0.055, a value consistent with measured values for similar disks in the literature and consistent with modeled vertical structure assuming the effects of radiation pressure and collisions} \citep{esposito2018,millar2015beta,krist2005hubble,thebault2009vertical}. The surface density is described by a smoothed broken/two-component power law, with two different fixed power law indices: one for the disk interior to a critical radius ($R_\text{c}$), set to $\alpha_\text{in}=-3.5$, and one for the disk exterior to $R_\text{c}$, set to $\alpha_\text{out}=1.5$ as in \citet{esposito2020debris}\rev{; these values were manually tuned to a reasonable by-eye match in that work, and we have adopted the same parameter choice for consistency.} The following model parameters were left free to vary in the fit: the minimum grain size ($a_\text{min}$), grain porosity, dust mass (within the grain size distribution used) $M_\text{d}$, disk position angle (PA), inclination (\textit{i}), dust inner radius $R_\text{in}$, dust outer radius $R_\text{out}$, and the critical radius $R_\text{c}$ at which the inner and outer surface density profiles cross. \rev{Although the $\alpha$ power law exponents can theoretically be used to define a disk edge where the signal falls below the noise, $R_\text{in}$ and $R_\text{out}$ are included as free parameters to be able to define an edge that is more abrupt than that from a smoothly declining power law, e.g. from planet sculpting. Additionally,} we do not expect the dust mass $M_\text{d}$ to be physically meaningful\rev{; instead, it essentially functions as a scaling factor for the global surface brightness of the disk.} PA is defined as the angle counterclockwise from North to the projected major axis of the disk.

\begin{figure*}
    \centering
    \includegraphics[width=0.9\linewidth]{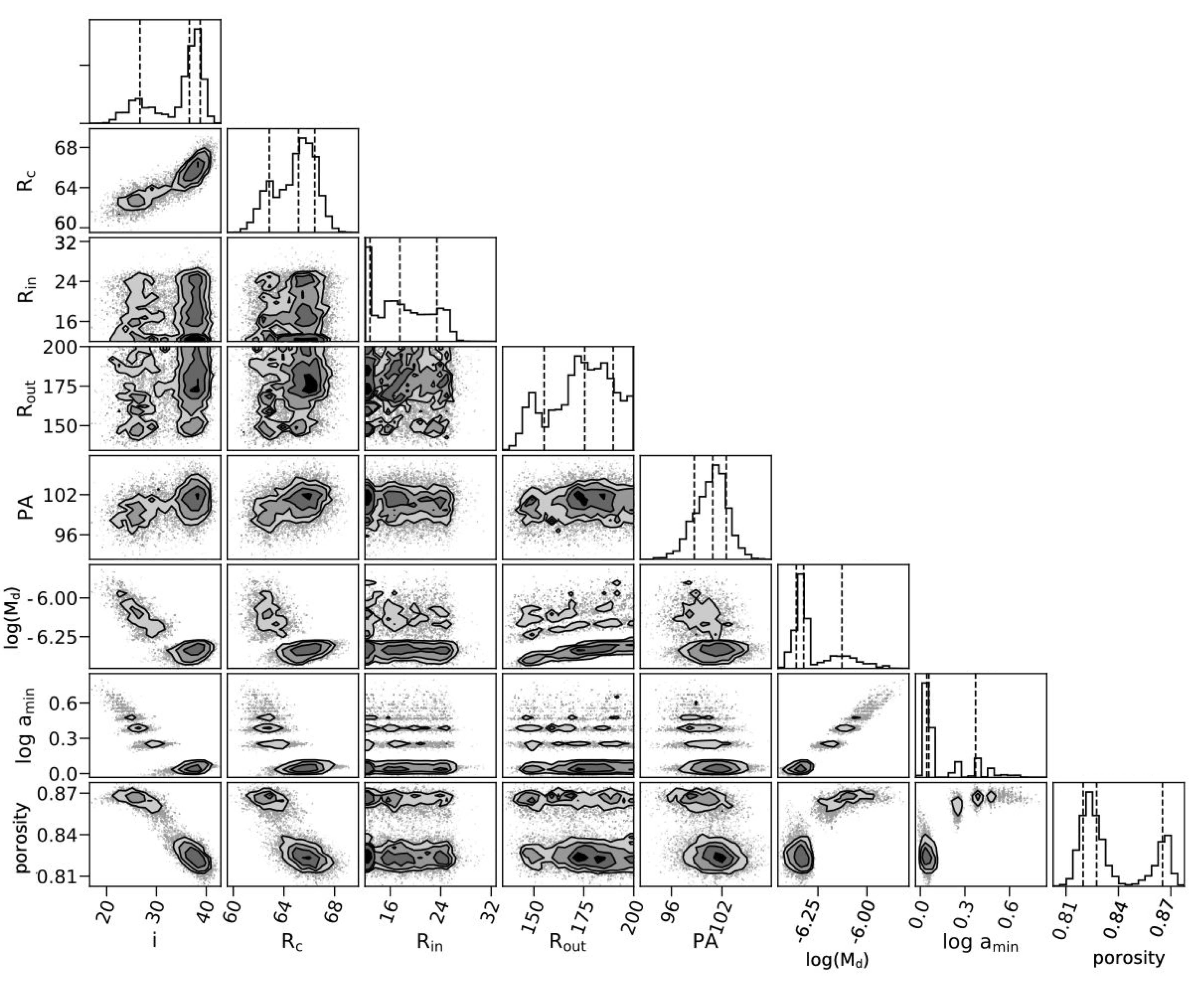}
    \caption{\rev{Corner plot of posteriors from \texttt{MCFOST} modeling of scattered light imaging.}}
    \label{mcfost-corner}
\end{figure*}

\begin{figure*}
    \centering
    \includegraphics[width=0.65\linewidth]{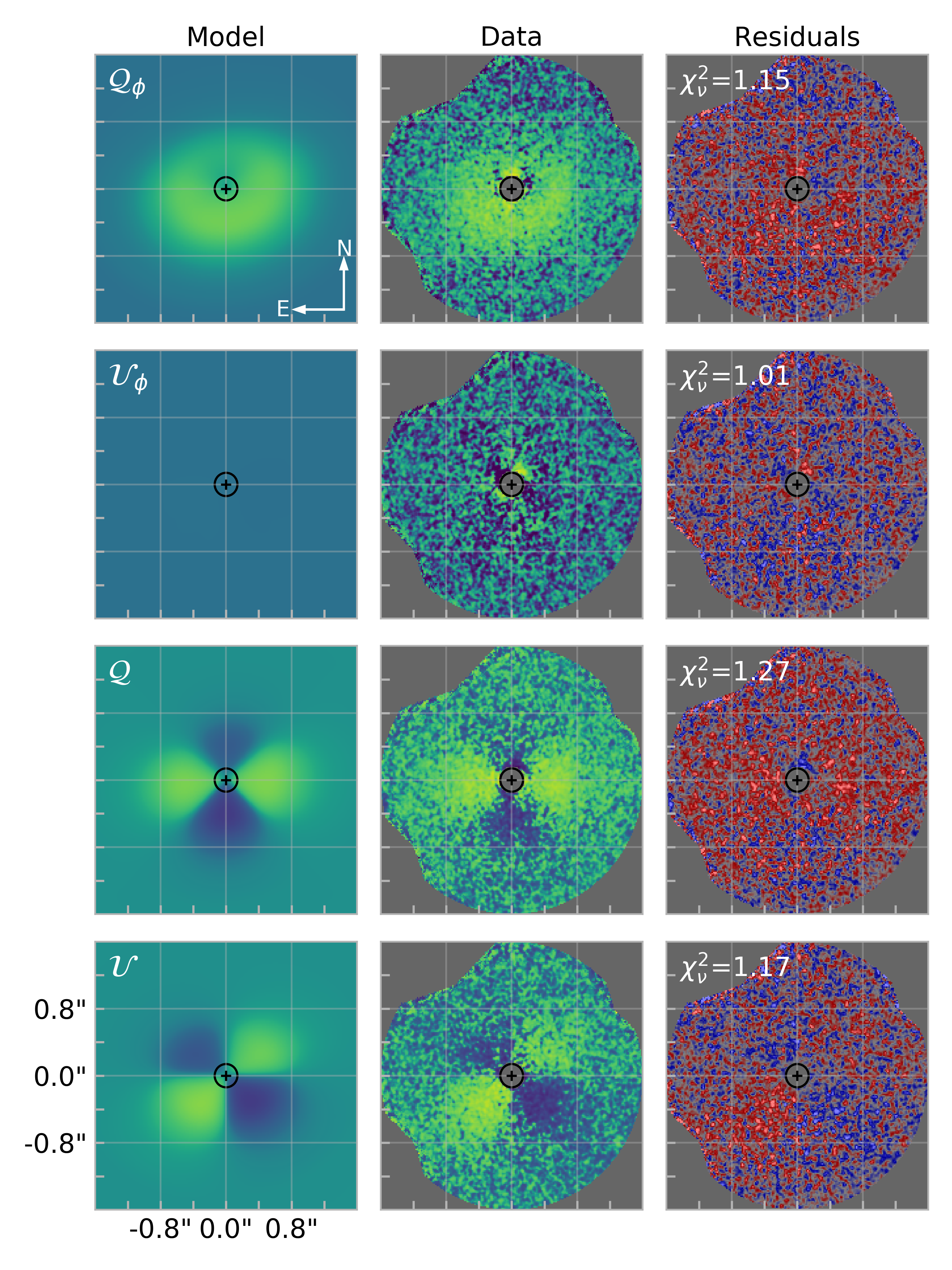}
    \caption{\rev{\texttt{MCFOST} maximum likelihood scattered light model of HD 156623 with data and residuals for Stokes $\mathcal{Q}_\phi$, $\mathcal{U}_\phi$, $\mathcal{Q}$, and $\mathcal{U}$, drawn from only the last 600 iterations (i.e. excluding burn-in).}}
    \label{fig:MCFOST}
\end{figure*}

We used a parallel-tempered MCMC with three temperatures, each with 120 walkers. Walkers were initialized randomly from a uniform distribution and then simulated for 3500 iterations. We discarded the first 2900 iterations as ``burn-in'' where the ensemble of walkers had not yet converged to their peak in posterior space, and used the final 600 iterations of the zeroth temperature walkers (equivalent to 2.16$\times 10^5$ models) for the disk parameter value estimates, which are summarized in Table \ref{mcfostparam}. ``Max Lk Value'' contains the values for the maximum likelihood model, which is shown in Figure \ref{fig:MCFOST} and the adjacent column shows the median values with a 1-sigma confidence interval. \rev{A corner plot of posteriors from the MCMC is shown in Figure \ref{mcfost-corner}. Multiple parameters have slightly multimodal or non-Gaussian posteriors, and as such the 1$\sigma$ values should be interpreted with caution for those parameters; parameters with strongly non-Gaussian posteriors are reported as either maximum likelihood values or 99.7\% confidence upper/lower limits.} Similar to \citet{esposito2020debris}, we aim for a model that constrains the necessary morphological parameters with sufficient accuracy for our purposes, while acknowledging that a more complex / physically-motivated treatment of grain properties is necessary to truly reproduce the observed scattered light distribution.

\begin{table*}
\resizebox{\textwidth}{!}{%
\begin{tabular}{|l|c|c|c|}
\multicolumn{4}{c}{\textbf{Best-Fit Morphological Parameters}} \\ \hline
\textbf{Parameter} & \textbf{Max. Lk. Value} & \textbf{50 $\pm$ 1$\sigma$} & \textbf{Value from \citet{esposito2020debris}} \\ \hline
PA (deg) & 102.081 & $100.9^{+1.6}_{-2.2}$ & 100.9$^{+1.9}_{-2.2}$ \\ \hline
Inclination (deg) & 38.978 & $36.6^{+2.1}_{-9.9}$ & 34.9$^{+3.6}_{-9.5}$ \\ \hline
$R_\text{in}$ (AU) & 12.707 & \rev{$<$ 26.6} & $<$ 26.7 (99.7\% confidence) \\ \hline
$R_\text{out}$(AU) & 184.482 & \rev{$>$ 139.2} & $>$ 139.3 (99.7\% confidence) \\ \hline
$R_\text{c}$(AU) & 67.113 & $65.2^{+1.3}_{-2.3}$ & 64.4$\pm$1.8 \\ \hline
\multicolumn{4}{c}{\textbf{Other Parameters Used}} \\ \hline
\textbf{Parameter} & \textbf{Free/Fixed} & \textbf{Max Lk. or Fixed Value} & \textbf{Value from \citet{esposito2020debris}} \\ \hline
$a_\text{min}$ ($\mu$m) & Free & 1.13 & 1.19  \\ \hline
$M_\text{d}$ (log$M_\odot$) & Free & -6.31 & -6.31$^{+0.21}_{-0.05}$ \\ \hline
Porosity (\%) & Free & 0.82 & 0.84 \\ \hline
$\alpha_\text{in}$ & Fixed & 1.5 & 1.5 \\ \hline
$\alpha_\text{out}$ & Fixed & -3.5 & -3.5 \\ \hline
$\beta$ & Fixed & 1 & 1 \\ \hline
$H_0/r_H$ & Fixed & 0.055 & 0.055 \\ \hline
\end{tabular}\label{mcfostparam}
}\caption{Summary of retrieved disk parameters from \texttt{MCFOST} from the final 600 iterations (i.e. excluding burn-in). The top section outlines the best fit morphological parameters adopted for our analysis, alongside the values from \citet{esposito2020debris} which are generally in agreement. The bottom section describes parameters which are not necessarily physically meaningful in this application and should be interpreted with caution, but are included here for reproduciblity of our models.}
\end{table*}

\section{Brightness and Scattering}\label{sec:bright}

\subsection{Radial Brightness Profiles}

Radial brightness profiles allow for further investigation of both the inner clearing and the outer radius of the disk, e.g. determining when the observed data is consistent with zero disk signal, in addition to the above modeling with \texttt{MCFOST}. 

\begin{figure}
    \centering
    \includegraphics[width=\linewidth]{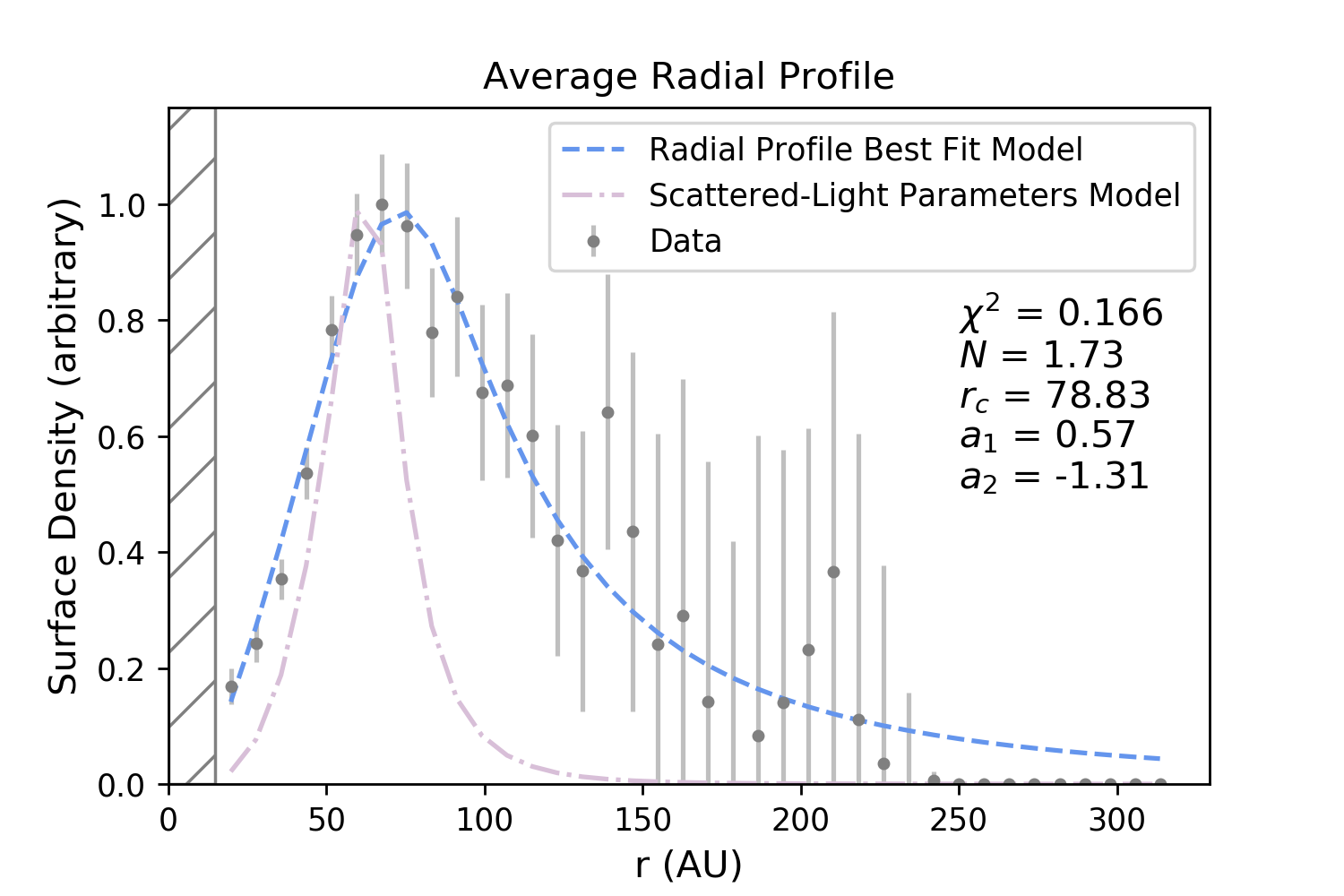}
    \caption{The average radial surface density profile (grey points with error bars) fit with a smoothly-broken two-component power law (the dashed blue line). The hatched region on the left marks the limits from the FPM as in other figures, and the dot-dash light purple curve represents the radial profile determined by scattered-light modeling. The values derived from this average surface density profile for $r_c$, $\alpha_\text{in}$, and $\alpha_\text{out}$ are substantially different from those determined via scattered-light modeling, perhaps due to a difference in the scattering phase function or bias due to the low SNR detection of the back-scattering portion of the disk\rev{; alternatively, these differences may simply be an artifact of the different modeling approaches.}}
    \label{fig:sdens-fit}
\end{figure}

\begin{figure*}
    \centering
    \includegraphics[width=0.49\linewidth]{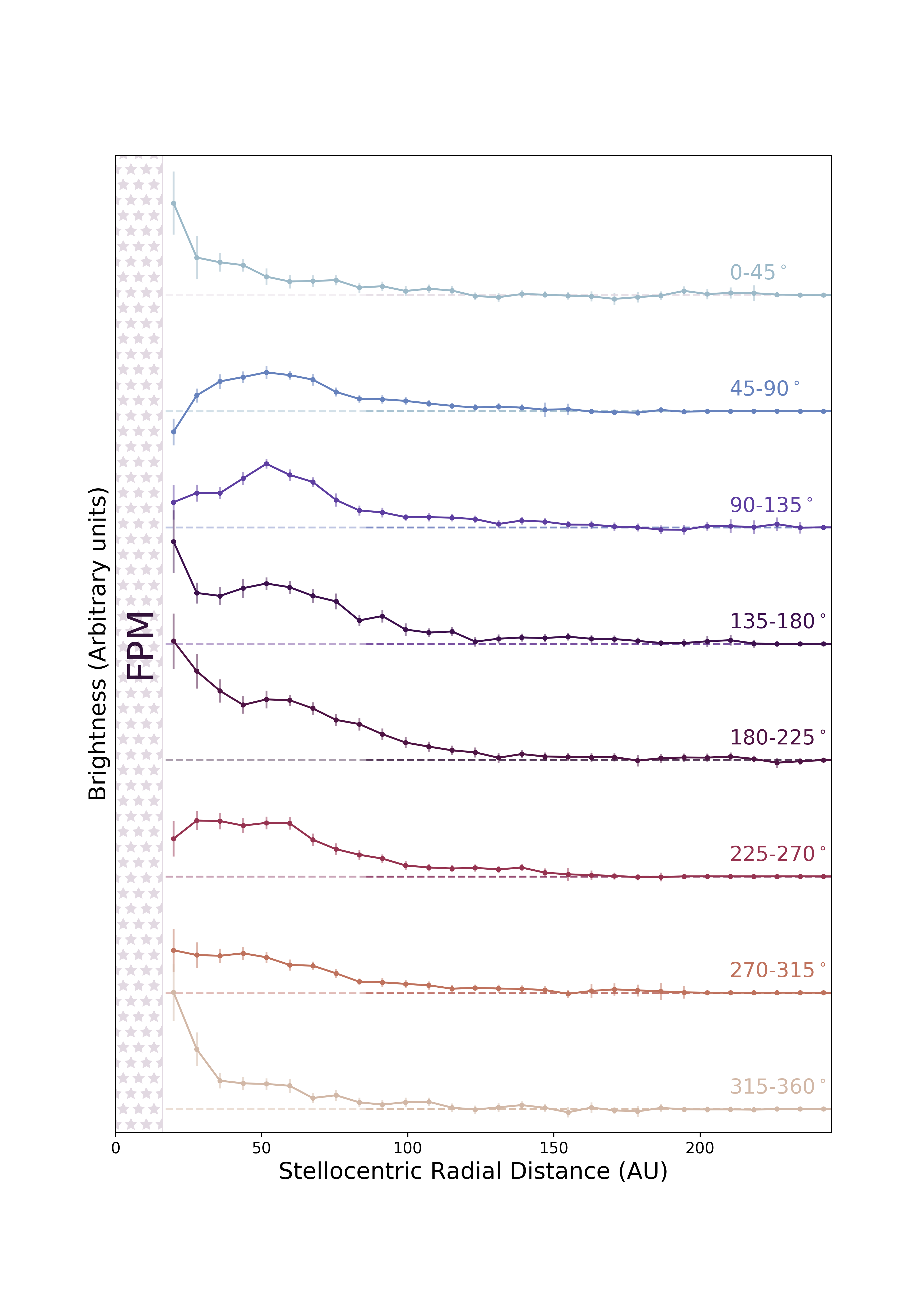}
    \includegraphics[width=0.49\linewidth]{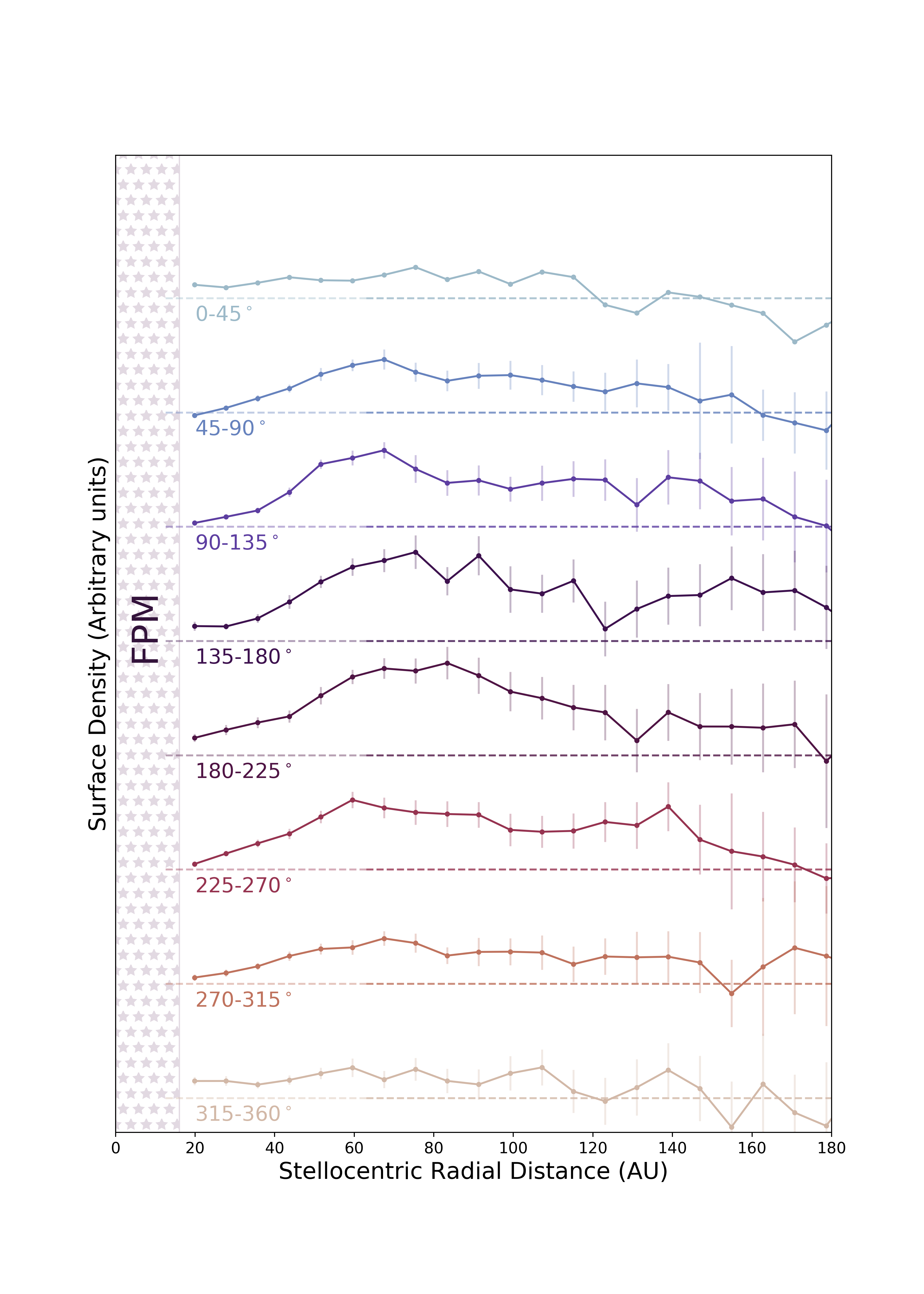}
    \caption{(Left) Measured brightness curves along the radial direction of HD 156623's deprojected disk for various angles. The region covered by the FPM is blocked out (hatched in light purple, labeled FPM), and it is clear that in most cases, \rev{the inner hole is unresolved, with the exception of possibly PA 45-90$^\circ$ and PA 90-135$^\circ$.} Each profile is arbitrarily offset for clarity, and the dashed lines correspond to zero for each profile depicted. (Right) Surface density (brightness corrected by a factor of 1/r$^2$) profiles along the radial direction of HD 156623's deprojected disk for various angles. The region covered by the FPM is blocked out (hatched in light purple, labeled FPM). Each profile is arbitrarily offset for clarity, and the dashed lines correspond to zero for each profile depicted.}
    \label{fig:brightcurv}
\end{figure*}

To that end, we measure radial brightness profiles (Figure \ref{fig:brightcurv}) for various position angles. To compute these profiles, we first deproject the disk based on our determined value of inclination, such that each radial cut traverses the same distance from the center of the disk, using the Python package \texttt{diskmap} \citep{2016A&A...596A..70S}. Zero position angle is defined counterclockwise from North, as described earlier. Radial brightness profiles are the average of a wedge at the given angle range that is \textbf{5} pixels long \rev{in the radial direction} \rev{and 45$^\circ$ ($\sim$5--80 pixels depending on the radial distance from the center) wide}, with its error as the variance of that slice of the wedge. Pixels within 15 AU of the center are masked, as this region is within the influence of the FPM (FPM) and therefore contains large errors. To determine a rough estimate for the visual outer radius of the disk--i.e. where the disk signal is dominant over the noise--in our data, we rebin the brightness profiles to 5 pixel wide annuli, and determine where the signal becomes consistent with zero by 2$\sigma$ in our images at each angle. We average those measurements across azimuthal angles to find a visual outer radius of $\sim$78 AU. This visual outer radius is not an inherent property of the disk, but instead a quantitatively-derived limit for where the image is dominated by the signal instead of the noise for use in further calculations, e.g. the scattering phase functions in Section \ref{sec:pol}. We additionally compute radial surface density profiles by correcting the brightness profiles by a factor of 1/$r^2$, as shown in Figure \ref{fig:sdens-fit}. As an independent check on the disk morphology, we fit a smoothly broken power law (as defined via MCFOST and below in Equation \ref{label:surfdens}) to the average measured surface density profile using least-squares, shown in Figure \ref{fig:sdens-fit}.

\begin{equation}\label{label:surfdens}
    \rho(r,z) \propto \frac{\exp[-|z|/H(r)^\gamma]}{[(r/r_c)^{-2\alpha_\text{in}} + (r/r_c)^{-2\alpha_\text{out}}]^{1/2}}
\end{equation}

We set the $z$ coordinate to a constant for simplicity, use the same formalism for $H(r) = H_0(r/r_H)^\beta$ as \citet{esposito2020debris} where $\beta$=0 and $H_0/r_H$=0.055, set the exponent $\gamma$=1, and leave $r_c$, $\alpha_\text{in}$, and $\alpha_\text{out}$ as free parameters, along with an arbitrary scaling/normalization factor $N$. The best fit values for the average radial surface density profile are \rev{$N$=1.73$\pm$0.13, $r_c$=78.8$\pm$6.0 AU, $\alpha_\text{in}$=0.58$\pm$0.15, and $\alpha_\text{out}$=-1.31$\pm$0.15.} These differ significantly from the scattered-light modeling values for $r_c$, $\alpha_\text{in}$, and $\alpha_\text{out}$. \rev{These differences may be explained simply by the fact that $\alpha_{\rm in}$ and $\alpha_{\rm out}$ were fixed in the \texttt{MCFOST} models, as $r_c$ is expected to vary with any change in the power law exponents.} Additionally, this average profile may be biased due to a lower SNR detection of the back-scattering half of the disk, where the outer edge is much less well-defined, or a different scattering phase function than that assumed by \texttt{MCFOST}, which may change the ratio of surface brightness to surface density.

\begin{figure}
    \centering
    \includegraphics[width=\linewidth]{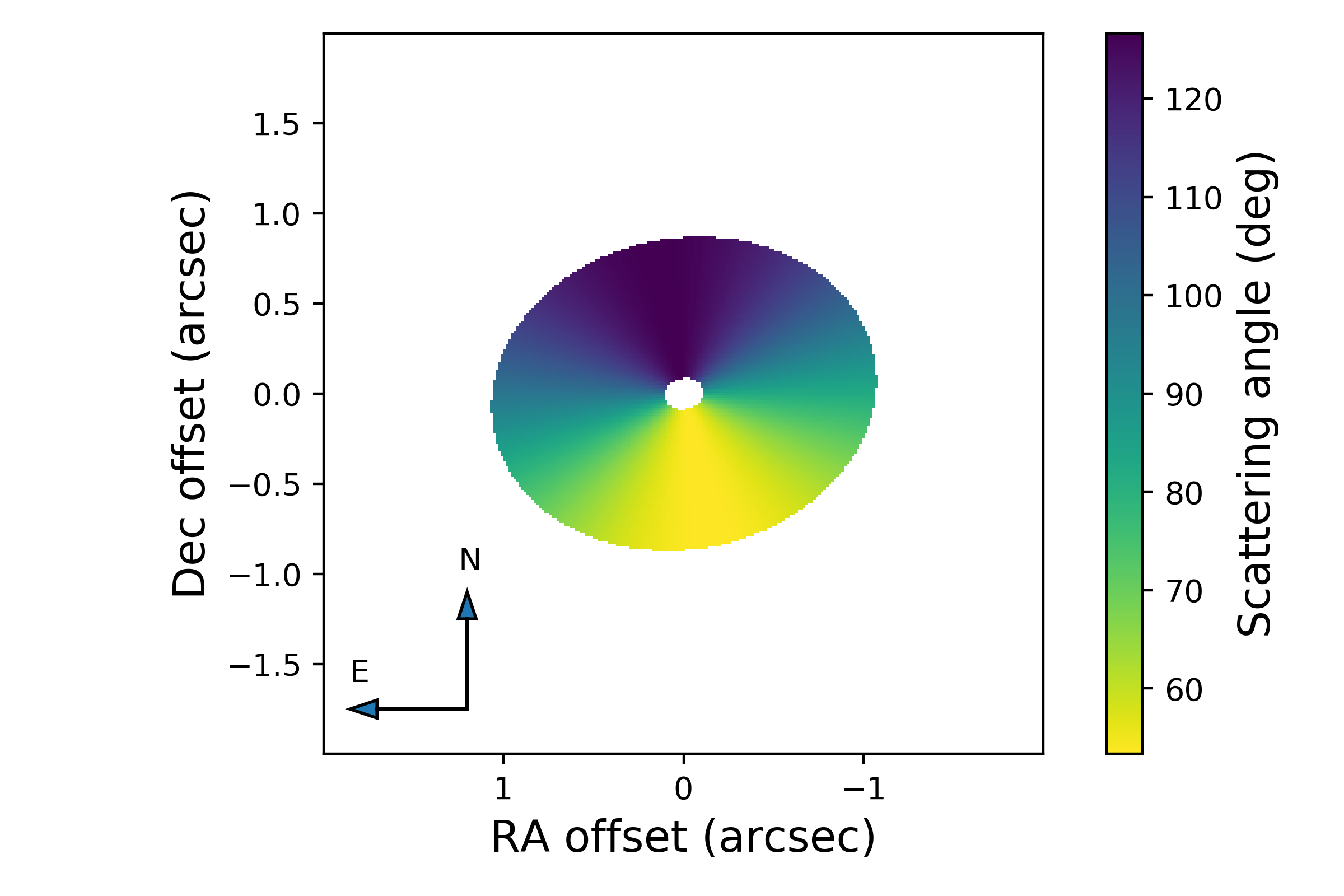}
    \caption{Diagram with the scattering angle across the disk of HD 156623, assuming the geometric parameters defined in Section \ref{subsec:MCFOST}. The most forward scattering occurs in the southwest, and the most back scattering occurs in the northeast.}
    \label{fig:phi_sca}
\end{figure}

\subsection{Observed Polarized Scattering Phase Functions} \label{sec:pol}

Scattering phase functions (SPFs) describe the fraction of light scattered into a solid angle by a single dust grain, and their shapes are determined by the properties of the dust itself; therefore, they are often used as a probe of the grain size, porosity, composition, and shape of a disk's dust grains. 

We make several assumptions in this calculation, similar to those used for geometric modeling; that is, we take the disk to be azimuthally symmetric (but radially varying, related to the observed surface brightness) and optically thin, with the previously-described morphological parameters inferred via \texttt{MCFOST}. We also assume that scattering is \rev{independent of azimuthal angle} (e.g. only the scattering angle $\theta$ is a variable in our equations), which is true for a spherical grain by symmetry and a good approximation for randomly oriented non-spherical grains. The expression for a normalized SPF is then as follows:
\begin{equation}
f(\theta) \sim \frac{B_\text{obs}(r,\theta)}{\rho(r)} \cdot r^2
\end{equation}
where $f(\theta)$ is the SPF, solely dependent on scattering angle and radially varying properties, $B_\text{obs}(r,\theta)$ is the observed brightness at a given radius $r$ and PA $\theta$, $\rho(r)$ is the surface density of the disk, derived from the observed radial brightness profile, and $r$ is the radius from the central star.

Given that this disk does not have a clearly defined single ring at which to compute a SPF, SPFs were instead determined by taking measurements of $B_\text{obs}(r,\theta)$ at multiple radii around the disk, then averaging over the varying radii. For this specific calculation, we limit the range over which we compute the SPF to the annuli beyond the FPM and within the visual outer radius as determined by the radial brightness curves; this range is 15 AU (at the FPM limit) to 78 AU. Since we do assume the disk's surface density can vary radially, we retain $\rho(r)$ and $r$ in our calculations, even though they are constants for a given radius. The disk surface density, $\rho(r)$, was approximated using the average best fit radial brightness profile where the peak surface brightness was normalized to 1.

Scattering angle ($\theta$) for each point in the disk is determined analytically, as shown in Figure \ref{fig:phi_sca}, following the above geometric assumptions. This disk is most forward scattering around $\sim$200$^\circ$ PA, with a scattering angle of 54$^\circ$, and most backward scattering near $\sim$10$^\circ$ PA with a scattering angle of 126$^\circ$.

\begin{figure*}
    \centering
    \includegraphics[width=0.45\linewidth]{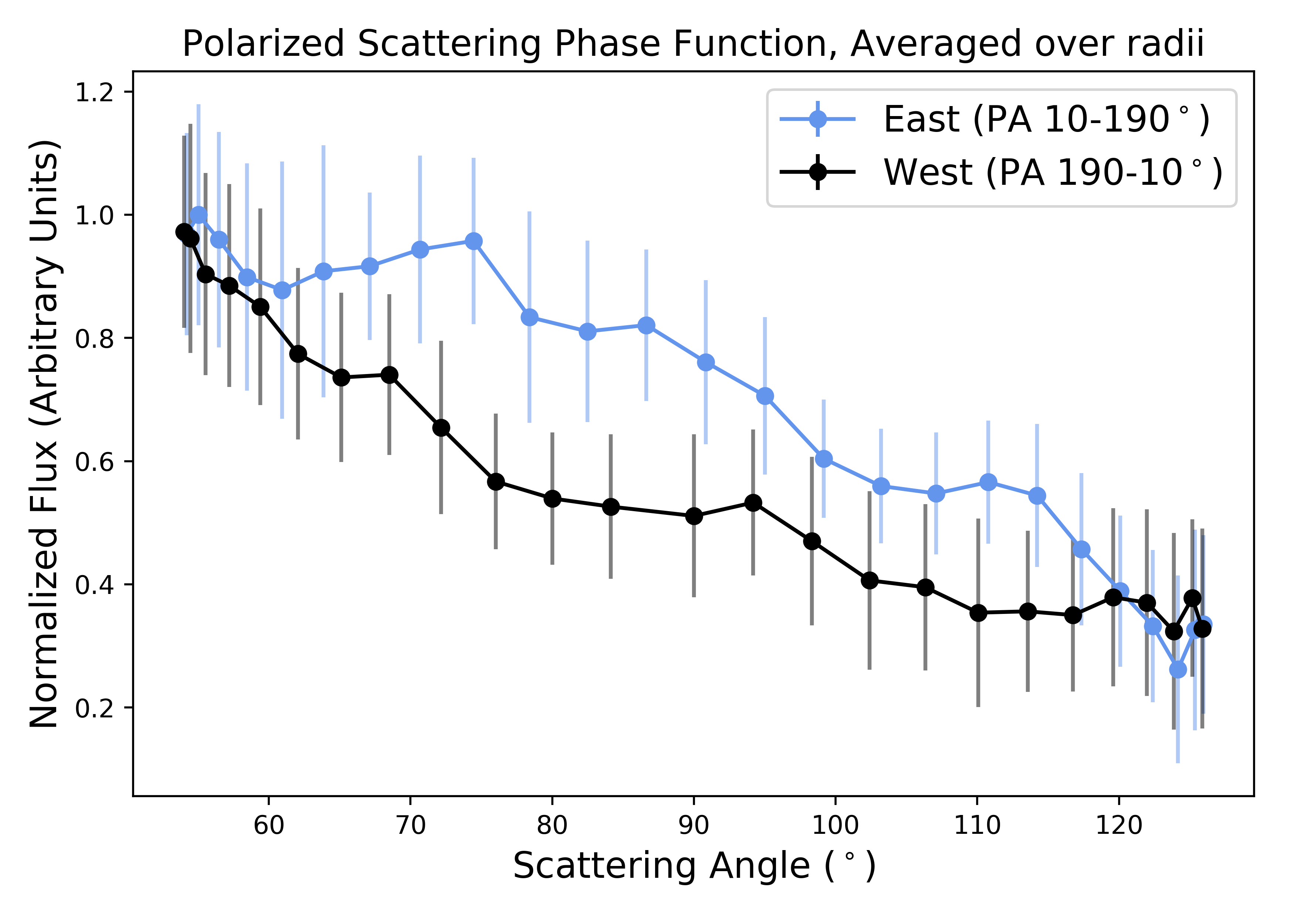}
    \includegraphics[width=0.45\linewidth]{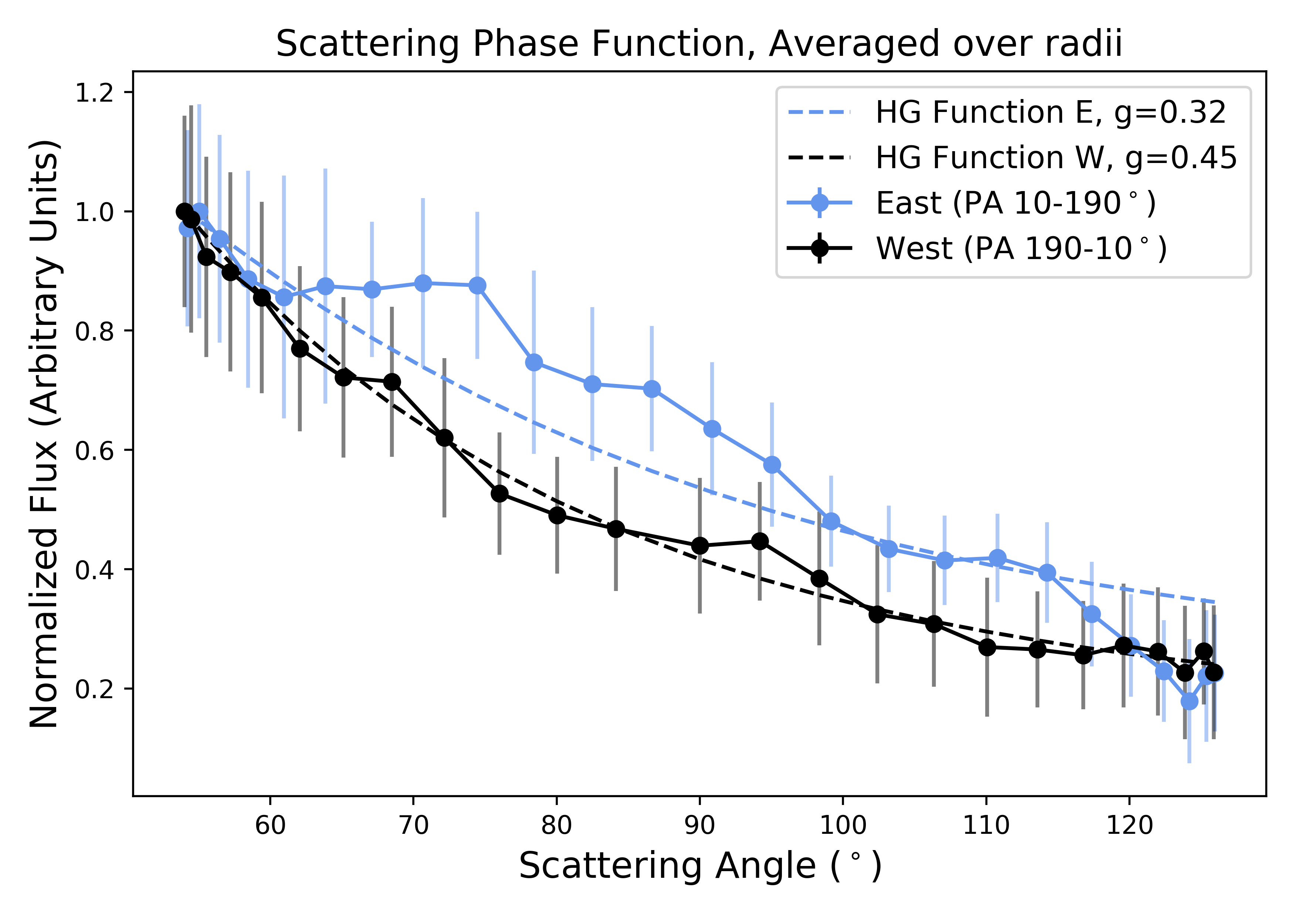}
    \caption{(Left) Measured polarized light scattering phase function for the disk, split between the Eastern and Western halves of the disk as defined by the approximate minor axis. We observe an asymmetry between the Eastern and Western halves, consistent with recent findings in \citet{crotts2024uniform} suggesting this disk may be slightly eccentric. (Right) The measured East/West SPFs divided by a Rayleigh polarization curve, along with Henyey-Greenstein functions fit to each SPF. There is a lesser but still distinct asymmetry between the two halves of the disk, and the Henyey-Greenstein function is a poor fit for the Eastern half.} 
    \label{fig:pfpol}
\end{figure*}

HD 156623's debris disk appears slightly asymmetric in visual inspection, particularly in the northwest quadrant, so SPFs were measured separately for the East and West halves of the disk (divided approximately along the minor axis), as shown in Figure \ref{fig:pfpol}. Errors were derived by the variance of pixels in each annulus, similar to the procedure for the radial brightness profiles, and propagated appropriately through the calculation. There is a slight asymmetry that is most pronounced near the eastern ansae ($\sim$80--95$^\circ$), consistent with the brightness asymmetry described in \citet{crotts2024uniform} that suggests the disk may be eccentric or experiencing more frequent collisions on the Eastern half. However, it is also worth noting that in our calculation, there is an implicit assumption that the density is azimuthally constant and radially varying. As a result, even if there is a true increase in brightness at those angles, there may not be a corresponding overbrightness in the SPF if we do not assume this symmetry.

We additionally fit a Henyey-Greenstein SPF \citep{henyey1941diffuse} to each of our observed scattering SPFs divided by a Rayleigh polarization curve (approximating the total intensity SPF), also shown in Figure \ref{fig:pfpol}. The Henyey-Greenstein (HG) function $p(\theta)$ is an empirical description, not based in Mie theory or another physical scenario, and it is calculated as 

\begin{equation}
    p(\theta) = \frac{1}{4\pi} \frac{1-g^2}{[1+g^2-2g\text{cos}\theta]^{3/2}}
\end{equation}

where $g$ is the Henyey-Greenstein asymmetry parameter ($g = \cos\theta$, ranging from -1 to 1) and $\theta$ is the scattering angle. For debris disks, which are typically forward scattering, $g$ is greater than 0. For the Eastern half of the disk, the HG function is a poor fit, even with the best fit value $g$ = 0.33; however, the HG function with $g$ = 0.45 is a reasonable approximation for the Western half. These $g$ values are broadly consistent with similar measurements for other low-inclination debris disks as summarized by \citet{hughes2018debris}, supporting the ideas that disk dust may not obey one Henyey-Greenstein SPF at all scattering angles and that best-fit $g$ values may correlate with the range of scattering angles probed. Although the HG function is not inherently physically informative, we do not pursue physical dust model fits to our phase function given the limited range of scattering angles for the available data. Additionally, there is still an asymmetry in the derived total intensity SPF--albeit less pronounced--again mirroring findings from \citet{crotts2024uniform}. 

\section{Thermal Emission} \label{sec:SED}

\begin{table*}
\begin{tabular}{|l|l|l|l|l|}
\hline
\textbf{Band} & \textbf{Source} & \textbf{$\lambda_\text{eff}$ ($\mu$m)} & \textbf{Flux (Jy)} & \textbf{References} \\ \hline
B & Tycho 2 & 0.44 & 4.75$\pm$0.10 & \citet{hog2000tycho,cox2015allen} \\ \hline
V & Tycho 2 & 0.55 & 4.71$\pm$0.1 & \citet{hog2000tycho,cox2015allen} \\ \hline
J & 2MASS & 1.24 & 2.33$\pm$0.05 & \citet{cohen2003spectral,skrutskie2006two} \\ \hline
\textit{H} & 2MASS & 1.66 & 1.47$\pm$0.03 & \citet{cohen2003spectral,skrutskie2006two} \\ \hline
K$_s$ & 2MASS & 2.16 & 1.05$\pm$0.02 & \citet{cohen2003spectral,skrutskie2006two} \\ \hline
W1 & ALLWISE & 3.35 & 0.502$\pm$0.027 & \citet{wright2010wide,mainzer2011preliminary,} \\ \hline
W2 & ALLWISE & 4.6 & 0.30 $\pm$0.010 & \citet{wright2010wide,mainzer2011preliminary} \\ \hline
S9W & \textit{AKARI} & 9 & 0.215$\pm$0.015 & \citet{murakami2007infrared,ishihara2010AKARI} \\ \hline
W3 & ALLWISE & 11.6 & 0.155$\pm$0.007 & \citet{wright2010wide,mainzer2011preliminary} \\ \hline
W4 & ALLWISE & 22.1 & 0.626$\pm$0.036 & \citet{wright2010wide,mainzer2011preliminary} \\ \hline
 & ALMA & 1240 & 0.00072$\pm$0.00011 & \citet{lieman2016debris} \\ \hline
\end{tabular}
\caption{Summary of photometry used in creating the spectral energy distribution for HD 156623.} \label{tab:sed}
\end{table*}

The main goal of our investigation of HD 156623's SED is to constrain the location of the disk's unseen inner edge. Photometric measurements of HD 156623 were acquired from a variety of sources, ranging from visible to mid-infrared to millimeter ALMA observations. Most measurements were obtained in magnitudes from the IRSA or Vizier databases, and then converted to Janskys using the zero points provided by each mission/project's documentation.  These observations and their references are summarized in Table \ref{tab:sed}; all data listed here are before color corrections, as we instead treat points with large bandpasses (namely W3/W4 and AKARI S9W) using synthetic photometry as described below. Although photometric data from \textit{IRAS} were available for this object, they were not used due to poor data quality flags present in the catalog. Given the wide bandpasses of the \textit{WISE} W3/W4 \citep{wright2010wide} and \textit{AKARI} S9W \citep{fujiwaraakari} filters, we compute synthetic photometry by integrating the model flux multiplied by the transmission profile for each filter (from \citet{rodrigo2020svo}) over its bandpass via the following equation as in \citet{tokunaga2005mauna}:
\begin{equation}
    \langle F_{\nu}\rangle = \frac{\int F_\nu(\nu) S(\nu)/\nu \; d\nu}{\int S(\nu)/\nu \; d\nu}
\end{equation}

where $\langle F_{\nu}\rangle$ is the monochromatic flux density (erg/s/cm$^2$/Hz) for our synthetic photometry, $F_\nu$ is the monochromatic flux density (erg/s/cm$^2$/Hz) from the object, $S(\nu)$ is the filter transmission function, and $\nu$ represents the frequencies in the bandpass. Additional absolute calibration uncertainties were incorporated for the WISE data points \citep{cutri2012vizier} and AKARI data points \citep{tanabe2008absolute}.

We initially fit a simple model consisting of two single-temperature blackbody curves for the stellar emission plus the disk's thermal emission to these monochromatic flux densities using a Levenberg-Marquardt least-squares fit implemented in \texttt{scipy}. The free parameters included are disk temperature ($T_d$) and a parameter proportional to the disk emitting area ($A$). The stellar radius ($R_*$) and temperature ($T_*$) are set using the values in Table \ref{tab:star} from \citet{mellon2019discovery}. The best-fit dust temperature is 184$\pm$23 K with a reduced chi squared $\chi_\nu^2$ value of 1.088, corresponding to a radius of 8.49$^{+2.60}_{-1.78}$ AU if treated as an equilibrium temperature for a grain radiating as a blackbody. \rev{It is additionally worth noting that radii derived from SEDs are known to be underestimates; i.e. they are often smaller than the inner radii derived from resolved observations \citep{wyatt2008evolution,pawellek2015dust,pawellek202175,booth2013resolved,hom2020first}.}

Given the poor fit at long wavelengths while using a single-temperature blackbody to represent thermal emission from dust, we subsequently tested a model for the spectral energy distribution using a single-temperature blackbody plus an additional modified wavelength-dependent emissivity, following the formalism in \citet{backman1992infrared}. Again, this model does not assume a spatial distribution or grain size distribution. In this model, grains of a characteristic size $\sim$$\lambda_0/2\pi$ radiate with a modified emissivity $\epsilon_\nu$ as follows:

\begin{equation}
  \epsilon_\nu =
    \begin{cases}
      \frac{\lambda_0}{\lambda} & \lambda>\lambda_0\\
      1 & \text{otherwise}
    \end{cases}       
\end{equation}

\begin{figure*}
    \centering
    \includegraphics[width=0.9\textwidth]{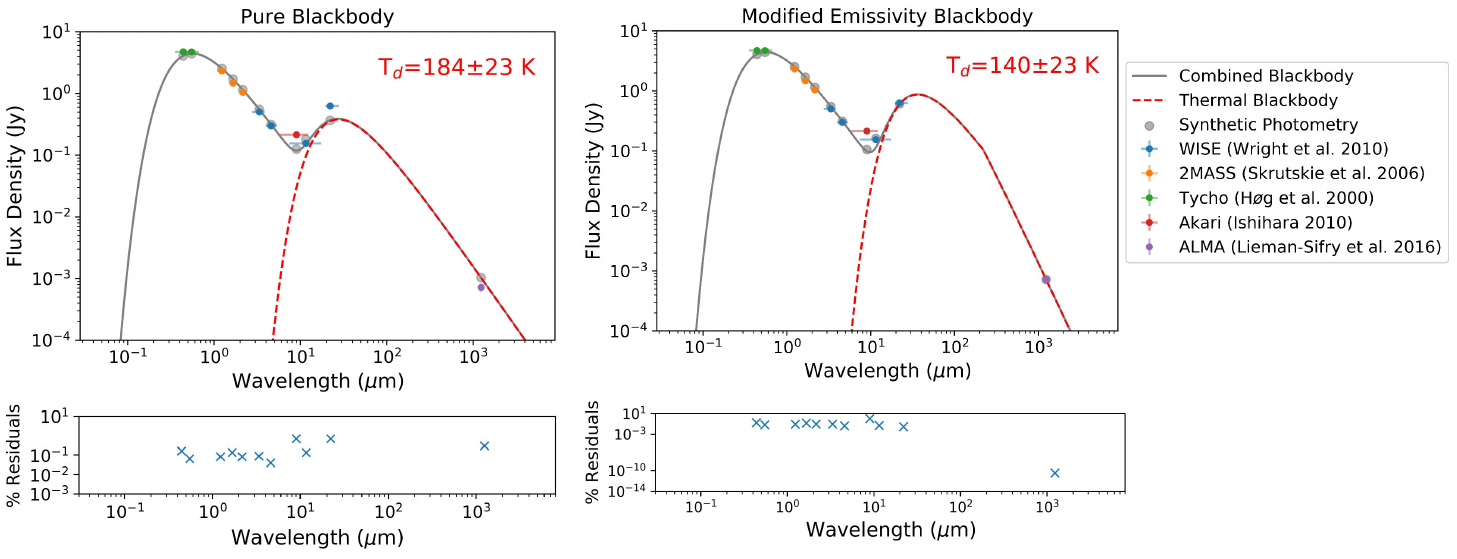}
    \caption{Single-component blackbody models using best-fit values to available photometry for HD 156623, with fractional residuals shown below. The dust's blackbody contribution to the SED is plotted in red. (Left) For a a simple single-component blackbody SED, the photometry is best fit with temperature T = 184$\pm$23 K. (Right) For a modified-emissivity single-component blackbody SED, the photometry is best fit with T = 140$\pm$23 K and $\lambda_0$ = 219 $\mu$m. Neither model is able to capture the 9$\mu$m AKARI point, and a modified emissivity is necessary to reproduce the flux observed by ALMA.}
    \label{fig:SED}
\end{figure*}

This model was again fit using the \texttt{scipy} implementation of a Levenberg-Marquardt least-squares fit, now for both temperature and $\lambda_0$. The best-fit disk temperature for this model is 140$\pm$23 K with a $\chi_\nu^2$ value of 1.21, corresponding to a radius of 14.7$^{+6.34}_{-3.85}$ AU, with $\lambda_0$ unconstrained (best fit value of 219 $\mu$m, corresponding to a characteristic grain size of $\sim$35 $\mu$m). These models are visualized in Figure \ref{fig:SED}. Neither model is able to capture the 9$\mu$m \textit{AKARI} point, and a modified emissivity is necessary to reproduce the flux observed by ALMA.


\begin{figure}
    \centering
    \includegraphics[width=\linewidth]{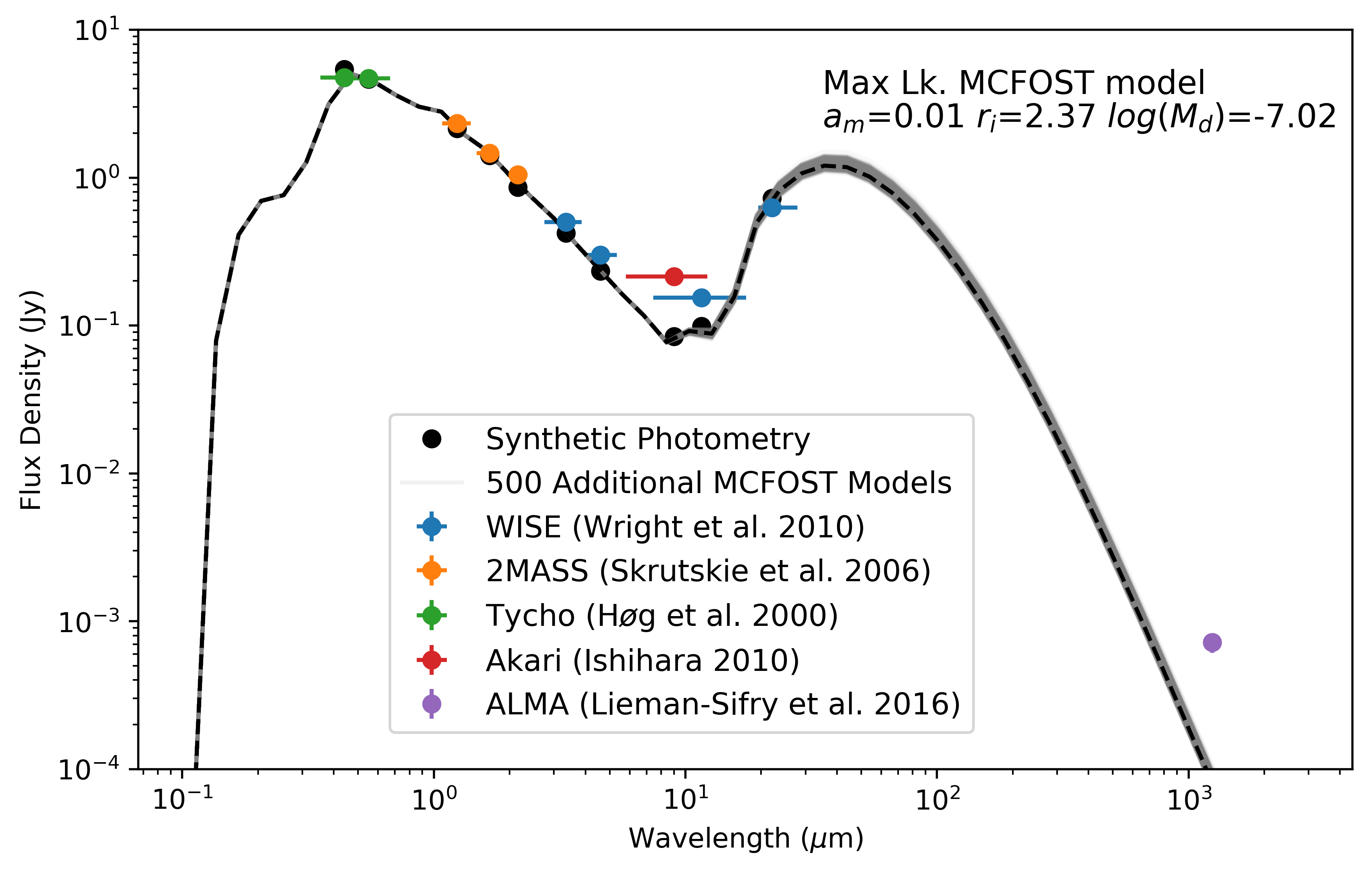}
    \caption{Maximum likelihood SED model from \texttt{MCFOST} (as a dashed black line, with synthetic photometric points marked) compared to available photometry (colored according to its source). The best fit model requires sub-blowout size grains and a close-in inner radius around $\sim$3 AU. 500 additional SED models randomly drawn from the post-burn-in MCMC are plotted in grey.
    }
    \label{fig:mcfost-sed}
\end{figure}

\begin{figure}
    \centering
    \includegraphics[width=\linewidth]{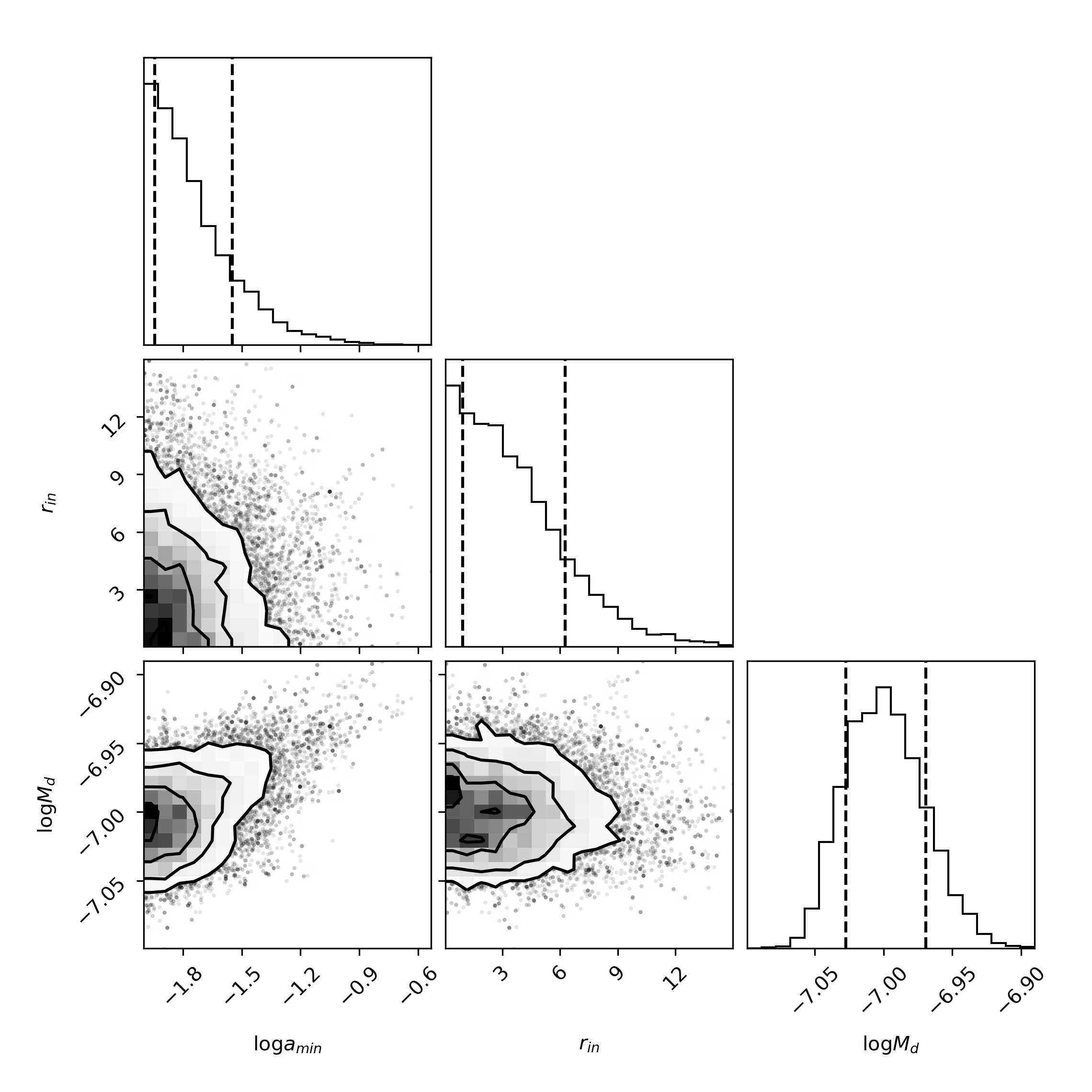}
    \caption{Corner plot for MCMC fit with photometry and \texttt{MCFOST} SED models. 
    \rev{The reported value for $r_{\rm in}$ is an upper bound ($<$ 13.43 AU at 99.7\% confidence), as seen in the posterior distribution. Additionally, the value for log$a_\text{min}$ is also only an upper limit of -0.685 (0.21 $\mu$m); }this suggests that different assumptions for the grain composition and/or size distribution may be needed to fully understand the minimum grain size of this system.}
    \label{fig:mcfost-sed-corner}
\end{figure}

However, these simple models do not explicitly fit for the inner radius we are interested in, as they assume a single temperature across the entire disk. To constrain the inner radius, we then compute the SED for a spatially resolved, azimuthally symmetric disk using \texttt{MCFOST}. We set up this resolved model using a Kurucz 9000\,K model atmosphere \citep{castelli2004new} and the same parameters and assumptions (e.g. Mie scattering, dust grain size and composition) as the scattered-light model in Section \ref{subsec:MCFOST}, leaving only $r_\text{in}$, $a_\text{min}$, and $M_\text{d}$ as free parameters. We explore the posterior for those parameters again with a \citet{goodman2010ensemble} affine invariant MCMC ensemble sampler using the \texttt{emcee} package with 16 walkers and uniform priors, with \rev{bounds on the priors for $r_{\rm in}$ (0$< r_{\rm in} <$15) and log$a_{\rm min}$ (-2$< {\rm log}a_{\rm min} <$1).} Walkers were initialized randomly from a uniform distribution and then simulated for 1500 iterations. We discarded the first 100 iterations as ``burn-in'' where the ensemble of walkers had not yet converged to their peak in posterior space, and used the remaining 1400 iterations (equivalent to 2.2$\times 10^4$ models) for the disk parameter value estimates. The maximum likelihood model SED is shown compared to available photometry in Figure \ref{fig:mcfost-sed}, and a corner plot of the approximate posteriors is shown in Figure \ref{fig:mcfost-sed-corner}.

Scattered-light image modeling is limited by the focal-plane mask in its determination of $r_\text{in}$, resulting in a maximum likelihood value that corresponds to the FPM edge ($\sim$12--15 AU) \rev{and a 3$\sigma$ upper limit of $<$26.6 AU; SED modeling, on the other hand, returns a maximum likelihood value of 2.37 AU and is limited by the prior bound at zero, resulting in an upper limit of $<$13.4 AU at 99.7\% confidence ($<$8.85 AU at 95\% confidence).} The value for log($M_\text{d}$) should be considered as a scaling factor more than a physically relevant value, as we assume a sparse grain density and do not consider dynamical effects like collisions; for the sake of reproducing our modeling, the value for log($M_\text{d}$) for the maximum-likelihood model shown in Figure \ref{fig:mcfost-sed} is -7.02. The minimum grain size $a_\text{min}$ is much smaller as suggested by SED modeling, \rev{$<$0.21 $\mu$m at compared to $\sim$1 $\mu$m from scattered-light image modeling. For comparison, the blowout size for the system is around 3.5 $\mu$m, as further discussed in Section \ref{sec:disc}. This $a_\text{min}$ value from SED modeling reached a bound on the prior at 0.01 $\mu$m,} suggesting that the Dohnanyi size distribution or other grain characteristics we have assumed may not fully describe this system; however, given the limited number of photometric data points currently available (four data points\rev{--AKARI 9$\mu$m, W3, W4, and ALMA--}and currently three free parameters), we choose not to add further free parameters to our model in this work. This best-fit MCFOST SED model has $\chi_\nu^2$ = 1.11, similar to the simpler models above. Again, the \textit{AKARI} and ALMA points are not fit well by this model; we discuss this discrepancy--as well as the discrepancy in $a_\text{min}$--further in Section \ref{sec:disc}. The maximum temperatures calculated by \texttt{MCFOST} at $r_\text{in}$ and $R_\text{c}$ at the disk midplane in the best fit model are $\sim$500 K and 355 K respectively.


\section{Discussion}\label{sec:disc}

All SED fits are unable to capture the \textit{AKARI} data point (the red point in Figure \ref{fig:SED}), and no single model we have tested is able to satisfactorily explain the entire set of photometric data points available. \texttt{MCFOST} modeling similarly has trouble reproducing the \textit{AKARI} point, as well as the long wavelength flux measured by ALMA, indicating that different assumptions for grain composition and/or size distribution may be needed; given the limited number of photometric points available, exploring those as free parameters in a model is beyond the scope of this work. However, based on our modeling, it appears that--under the assumptions described in our work--the thermal signature of the disk is best explained by a population of close-in warm dust with sub-blowout size grains, and a modified emissivity trend at longer wavelengths. One could additionally imagine that the debris disk has two components: a cold outer disk imaged by GPI and ALMA, and a separate warm inner belt that is behind the GPI FPM and unresolved by ALMA, but also boosts the 9 $\mu$m emission as seen by \textit{AKARI}. Future work may include exploring two-component models, and further observations (e.g. \textit{JWST}/MIRI spectroscopy) are necessary to better define the shape of the spectrum through the mid-infrared.  

The radiative transfer modeling of the scattered light data is limited by the focal-plane mask, initially constraining $r_\text{in}$ to be less than \rev{26.6} AU. Single-component blackbody fits to the SED suggest an inner edge of $\sim$8-15 AU (not accounting for non-blackbody behavior of real dust grains or the possibility of grain overheating; e.g. \citet{pawellek2015dust,morales2011common}), and disk models with \texttt{MCFOST} fit to the SED constrain the edge to \rev{$<$ 13.4} AU. However, it is again worth noting that even the best fit SED model is unable to reproduce both the 9 $\mu$m excess and the long wavelength flux observed by ALMA, indicating that different assumptions on the grain properties or surface density distribution (e.g. inner and outer power laws) are necessary to fully explain this system; given the lack of photometry at this time, we leave this endeavor for future work. 
\citet{lieman2016debris} similarly constrains the inner radius of this disk using ALMA data and SED modeling, with a best fit value of $r_{in}$ = 10 AU from a visibility-only fit and $r_{in}$ = 2.99 AU from a simultaneous visibility and SED fit. Our result is in agreement with these values within 1$\sigma$. This \rev{$<$ 13.4} AU inner edge is certainly unique; HD 156623 is one of few known low-inclination debris disks in Sco-Cen \rev{imaged in scattered light (which is biased towards edge-on disks)} where the inner clearing is not visible due to its location behind the FPM \citep{bonnefoy2021narrow}. Additionally, this modeling assumes a certain power law for surface density; different assumptions in that value may lead to significant changes in the value of $r_\text{in}$, possibly explaining the slight differences between our estimate and that of \citet{lieman2016debris}. 

The blowout grain size of this system is estimated as $\sim$3.5 $\mu$m, according to \citet{arnold2019effect} which uses the equation below \citep{burns1979radiation} to compute the blowout size for HR 4796, a similar temperature A-type star.
\begin{equation}
    a_\text{BO} = 1.152\langle Q_\text{pr} \rangle\left(\frac{L_*}{L_\odot}\right)\left(\frac{M_*}{M_\odot}\right)^{-1}(1-\mathcal{P})\left(\frac{\rho}{\text{g/cm}^3}\right)^{-1}
\end{equation}
where $L_*$ is the stellar luminosity is solar luminosities, $M_*$ is the stellar mass in solar masses, $\mathcal{P}$ is the grain porosity (assumed to be zero), $\rho$ is the grain density (assumed to be 3.3 g/cm$^3$ from \citet{Draine1984ApJ,draine2003interstellar}), and $\langle Q_\text{pr}\rangle$ is the average radiation pressure efficiency, calculated from grain scattering properties through:
\begin{equation}
    \langle Q_{\mathrm{pr}}\rangle = \frac{1}{n} \sum_{i={\nu_{\mathrm{min}}}}^{\nu_{\mathrm{max}}} [F_{\nu,*}Q_{\mathrm{abs,i}} + (1-g)Q_{\mathrm{sca,i}}].
\end{equation}  
where $\nu_\text{min}$ is the lowest frequency used in the calculation, $\nu_\text{max}$ is the highest frequency used in the calculation, $F_{\nu,*}$ is the stellar spectrum at a given frequency, $Q_\text{abs,i}$ is the absorption coefficient for a given frequency, $g$ is the asymmetry parameter, $Q_\text{sca,i}$ is the scattering coefficient for a given frequency, and $n$ is the number of frequency samples used. 
We will adopt the blowout size of 3.5 $\mu$m for the remainder of this discussion which assumes astrosilicate grains as we have in our modeling.

Values for the minimum grain size from both scattered-light image modeling and SED modeling are in significant disagreement with the calculated blowout size; additionally, these values constrain the minimum grain size from different data and different assumptions on the physics of grain interactions with light, as the image modeling value is derived from the polarized scattering phase function and the SED modeling value is derived from thermal equilibrium. \texttt{MCFOST} scattered-light image modeling reports the maximum likelihood value for $a_{min}$ as 1.13 $\mu$m, with smaller peaks in the probability distribution around 1.35--1.82 $\mu$m \rev{as shown in Figure \ref{mcfost-corner}}, and SED modeling suggests $< 1 \mu$m; however, these values are much smaller than the minimum dust grain size predicted by the blowout size of 3.5 microns. \citet{lieman2016debris} provides additional evidence for sub-blowout size grains via their simultaneous model fits to their observed ALMA visibilities and the SED, finding a minimum grain size around $\sim$1 $\mu$m, in agreement with our scattered-light modeling. 
This discrepancy from the blowout size could indicate a rapid re-supply of small grains, possibly from ongoing collisions; however, this blowout size could alternatively indicate that we have underestimated the minimum grain size from radiative transfer and grain modeling due to our assumptions on grain properties \citep{grigorieva2007collisional} and/or the use of Mie theory, which is known to be a poor representation of true grain properties. However, if the grains are, in fact, porous as indicated by our radiative transfer modeling, the blowout size could be significantly larger, making this discrepancy even larger. The blowout timescale is approximately the free-fall time
, which is short, on the order of orbital timescales. As a result, the blowout timescale should be the dominant timescale for this system. The collision timescale, which dominates the dust replenishment, 
is $\sim 10^2$ yr--substantially longer than the blowout timescale. 

As HD 156623 is known to host a gas-rich disk, gas drag could change the blowout size and associated timescales--that is, small dust grains may be coupled to the gas, enabling their retention and explaining the observed inner edge and sub-blowout size grains. In \citet{lieman2016debris}, HD 156623 has one of the strongest CO \textit{J} = 2-1 detections out of their sample in Sco-Cen; perhaps this system has a particularly high gas mass, and therefore greater influence of gas drag on the dust dynamics. The sample of Sco-Cen disks from \citet{lieman2016debris} also suggests a correlation between the presence of gas and sub-blowout grains, as seen in this system. Further modeling is needed to fully understand the impact of the presence of gas in this disk.

\section{Conclusions} \label{sec:conclusions}

We have (1) presented analysis of the morphology of the possible ``hybrid'' debris disk around HD 156623 in \textit{H} band polarized scattered light, (2) presented an empirical measurement of the brightness/surface density profile and scattering phase function for HD 156623's debris disk, and 3) determined that the inner radius of the disk is likely close to the star to \rev{$<$ 13.4 AU} via SED modeling and imaging. Our work suggests that HD 156623 retains sub-blowout size grains, and gas drag may be a key influence on the dust population of this debris disk, in line with findings from \citet{lieman2016debris}. However, we acknowledge the numerous degeneracies present in this type of debris disk data, as the observed scattered-light distribution is a complicated and difficult-to-disentangle combination of disk geometry, grain size distribution, grain properties (e.g. porosity, scattering efficiency), and more \citep{stark2014revealing}. Future work should consider obtaining additional mid-infrared photometry (e.g. from \textit{JWST}) to further constrain the debris disk SED, and testing more complex models, such as a two-component dust population or simultaneously modeling the gas and dust components of the disk to explore how gas drag impacts small grain retention in this system. %

\section*{Acknowledgements}

This material is based upon work supported by the National Science Foundation Graduate Research Fellowship under Grant No. 2016-21 DGE-1650604 and 2021-25 DGE-2034835. Any opinions, findings, and conclusions or recommendations expressed in this material are those of the authors(s) and do not necessarily reflect the views of the National Science Foundation. KAC and BCM acknowledge a Discovery Grant from the Natural Science and Engineering Research Council of Canada.

This work is based on data obtained by the international Gemini Observatory, a program of NSF’s NOIRLab, which is managed by the Association of Universities for Research in Astronomy (AURA) under a cooperative agreement with the National Science Foundation on behalf of the Gemini Observatory partnership: the National Science Foundation (United States), National Research Council (Canada), Agencia Nacional de Investigaci\'{o}n y Desarrollo (Chile), Ministerio de Ciencia, Tecnolog\'{i}a e Innovaci\'{o}n (Argentina), Minist\'{e}rio da Ci\^{e}ncia, Tecnologia, Inova\c{c}\~{o}es e Comunica\c{c}\~{o}es (Brazil), and Korea Astronomy and Space Science Institute (Republic of Korea).

This publication makes use of data from the following:
(1) the Wide-field Infrared Survey Explorer, which is a joint project of the University of California, Los Angeles, and the Jet Propulsion Laboratory/California Institute of Technology, and NEOWISE, which is a project of the Jet Propulsion Laboratory/California Institute of Technology. \textit{WISE} and NEOWISE are funded by the National Aeronautics and Space Administration (2) the Two Micron All Sky Survey, which is a joint project of the University of Massachusetts and the Infrared Processing and Analysis Center/California Institute of Technology, funded by the National Aeronautics and Space Administration and the National Science Foundation (3) the European Space Agency (ESA) mission {\it Gaia} (\url{https://www.cosmos.esa.int/gaia}), processed by the {\it Gaia} Data Processing and Analysis Consortium (DPAC, \url{https://www.cosmos.esa.int/web/gaia/dpac/consortium}). Funding for the DPAC has been provided by national institutions, in particular the institutions participating in the {\it Gaia} Multilateral Agreement. This research is based on observations with \textit{AKARI}, a JAXA project with the participation of ESA. The Infrared Astronomical Satellite (\textit{IRAS}) was a joint project of the US, UK and the Netherlands (4) ADS/JAO.ALMA\#2012.1.00688.S. ALMA is a partnership of ESO (representing its member states), NSF (USA) and NINS (Japan), together with NRC (Canada), MOST and ASIAA (Taiwan), and KASI (Republic of Korea), in cooperation with the Republic of Chile. The Joint ALMA Observatory is operated by ESO, AUI/NRAO and NAOJ. The National Radio Astronomy Observatory is a facility of the National Science Foundation operated under cooperative agreement by Associated Universities, Inc.

This research has made use of NASA's Astrophysics Data System Bibliographic Services, the VizieR catalogue access tool, CDS, Strasbourg, France, and, for the following data sets, the NASA/IPAC Infrared Science Archive, which is funded by the National Aeronautics and Space Administration and operated by the California Institute of Technology: \citet{cutri20032mass,neugebauer2019IRAS,wright2019allWISE,AKARI2020AKARI}. This research has made use of the Spanish Virtual Observatory (https://svo.cab.inta-csic.es) project funded by MCIN/AEI/10.13039/501100011033/ through grant PID2020-112949GB-I00.

This work also made use of Astropy: a community-developed core Python package and an ecosystem of tools and resources for astronomy \citep{astropy:2013, astropy:2018, astropy:2022} and the \texttt{imgmasks} package, developed by B. Lewis, R. Lopez, E. Teng, and J. Williams as part of the 2021 Code / Astro workshop. Thanks to Elizabeth and Jason for their work on this package, and to the Code / Astro leads (namely Sarah Blunt and Jason Wang) for a great workshop.

The authors would also like to thank Dr. A. Meredith Hughes and Dr. John Carpenter for their helpful correspondence regarding the data used in Lieman-Sifry et al. 2016, Marcos M. Flores for showing B. Lewis that \texttt{np.arctan2} exists, Quinn Pratt and Joseph Marcinik for helpful discussions on interpolation and numerical integration, and Minghan Chen for helpful discussions on fitting for disk geometry. Thanks to Carlos Quiroz and Fredrik Rantakyro for assistance in obtaining seeing monitor records from Gemini Observatory for our data from 2019. B. Lewis would also like to acknowledge the invaluable labor of the maintenance and clerical staff at her institution and the observatories, whose contributions make scientific discovery a reality. This research was partially conducted on the traditional lands of the Gabrielino-Tongva people. 

\software{NumPy \citep{numpy},
IPython \citep{ipython}, Jupyter Notebooks \citep{jupyter},
Matplotlib \citep{matplotlib}, 
Astropy \citep{astropy:2013, astropy:2018, astropy:2022}, SciPy \citep{scipy}, pyKLIP \citep{wang2015pyklip}, MCFOST \citep{pinte2006monte,pinte2009benchmark}, pymcfost \citep{pymcfost}, diskmc \citep{esposito2018}, diskmap \citep{2016A&A...596A..70S}, imgmasks \citep{briley_l_lewis_2021_5029487}, emcee \citep{foreman2013emcee,foreman2019emcee},mcfost-python \citep{wolff2017hubble,mcpy}}

\bibliography{disk,atc}

\begin{thebibliography}{}
\expandafter\ifx\csname natexlab\endcsname\relax\def\natexlab#1{#1}\fi
\providecommand{\url}[1]{\href{#1}{#1}}
\providecommand{\dodoi}[1]{doi:~\href{http://doi.org/#1}{\nolinkurl{#1}}}
\providecommand{\doeprint}[1]{\href{http://ascl.net/#1}{\nolinkurl{http://ascl.net/#1}}}
\providecommand{\doarXiv}[1]{\href{https://arxiv.org/abs/#1}{\nolinkurl{https://arxiv.org/abs/#1}}}

\bibitem[{AKARI(2020)}]{AKARI2020AKARI}
AKARI, T. 2020, IPAC, doi, 10, 26131, \dodoi{10.26131/IRSA181}

\bibitem[{Arnold {et~al.}(2019)Arnold, Weinberger, Videen, \&
  Zubko}]{arnold2019effect}
Arnold, J.~A., Weinberger, A.~J., Videen, G., \& Zubko, E.~S. 2019, The
  Astronomical Journal, 157, 157

\bibitem[{{Astropy Collaboration} {et~al.}(2013){Astropy Collaboration},
  {Robitaille}, {Tollerud}, {Greenfield}, {Droettboom}, {Bray}, {Aldcroft},
  {Davis}, {Ginsburg}, {Price-Whelan}, {Kerzendorf}, {Conley}, {Crighton},
  {Barbary}, {Muna}, {Ferguson}, {Grollier}, {Parikh}, {Nair}, {Unther},
  {Deil}, {Woillez}, {Conseil}, {Kramer}, {Turner}, {Singer}, {Fox}, {Weaver},
  {Zabalza}, {Edwards}, {Azalee Bostroem}, {Burke}, {Casey}, {Crawford},
  {Dencheva}, {Ely}, {Jenness}, {Labrie}, {Lim}, {Pierfederici}, {Pontzen},
  {Ptak}, {Refsdal}, {Servillat}, \& {Streicher}}]{astropy:2013}
{Astropy Collaboration}, {Robitaille}, T.~P., {Tollerud}, E.~J., {et~al.} 2013,
  \aap, 558, A33, \dodoi{10.1051/0004-6361/201322068}

\bibitem[{{Astropy Collaboration} {et~al.}(2018){Astropy Collaboration},
  {Price-Whelan}, {Sip{\H{o}}cz}, {G{\"u}nther}, {Lim}, {Crawford}, {Conseil},
  {Shupe}, {Craig}, {Dencheva}, {Ginsburg}, {Vand erPlas}, {Bradley},
  {P{\'e}rez-Su{\'a}rez}, {de Val-Borro}, {Aldcroft}, {Cruz}, {Robitaille},
  {Tollerud}, {Ardelean}, {Babej}, {Bach}, {Bachetti}, {Bakanov}, {Bamford},
  {Barentsen}, {Barmby}, {Baumbach}, {Berry}, {Biscani}, {Boquien}, {Bostroem},
  {Bouma}, {Brammer}, {Bray}, {Breytenbach}, {Buddelmeijer}, {Burke},
  {Calderone}, {Cano Rodr{\'\i}guez}, {Cara}, {Cardoso}, {Cheedella}, {Copin},
  {Corrales}, {Crichton}, {D'Avella}, {Deil}, {Depagne}, {Dietrich}, {Donath},
  {Droettboom}, {Earl}, {Erben}, {Fabbro}, {Ferreira}, {Finethy}, {Fox},
  {Garrison}, {Gibbons}, {Goldstein}, {Gommers}, {Greco}, {Greenfield},
  {Groener}, {Grollier}, {Hagen}, {Hirst}, {Homeier}, {Horton}, {Hosseinzadeh},
  {Hu}, {Hunkeler}, {Ivezi{\'c}}, {Jain}, {Jenness}, {Kanarek}, {Kendrew},
  {Kern}, {Kerzendorf}, {Khvalko}, {King}, {Kirkby}, {Kulkarni}, {Kumar},
  {Lee}, {Lenz}, {Littlefair}, {Ma}, {Macleod}, {Mastropietro}, {McCully},
  {Montagnac}, {Morris}, {Mueller}, {Mumford}, {Muna}, {Murphy}, {Nelson},
  {Nguyen}, {Ninan}, {N{\"o}the}, {Ogaz}, {Oh}, {Parejko}, {Parley}, {Pascual},
  {Patil}, {Patil}, {Plunkett}, {Prochaska}, {Rastogi}, {Reddy Janga},
  {Sabater}, {Sakurikar}, {Seifert}, {Sherbert}, {Sherwood-Taylor}, {Shih},
  {Sick}, {Silbiger}, {Singanamalla}, {Singer}, {Sladen}, {Sooley},
  {Sornarajah}, {Streicher}, {Teuben}, {Thomas}, {Tremblay}, {Turner},
  {Terr{\'o}n}, {van Kerkwijk}, {de la Vega}, {Watkins}, {Weaver}, {Whitmore},
  {Woillez}, {Zabalza}, \& {Astropy Contributors}}]{astropy:2018}
{Astropy Collaboration}, {Price-Whelan}, A.~M., {Sip{\H{o}}cz}, B.~M., {et~al.}
  2018, \aj, 156, 123, \dodoi{10.3847/1538-3881/aabc4f}

\bibitem[{{Astropy Collaboration} {et~al.}(2022){Astropy Collaboration},
  {Price-Whelan}, {Lim}, {Earl}, {Starkman}, {Bradley}, {Shupe}, {Patil},
  {Corrales}, {Brasseur}, {N{"o}the}, {Donath}, {Tollerud}, {Morris},
  {Ginsburg}, {Vaher}, {Weaver}, {Tocknell}, {Jamieson}, {van Kerkwijk},
  {Robitaille}, {Merry}, {Bachetti}, {G{"u}nther}, {Aldcroft},
  {Alvarado-Montes}, {Archibald}, {B{'o}di}, {Bapat}, {Barentsen}, {Baz{'a}n},
  {Biswas}, {Boquien}, {Burke}, {Cara}, {Cara}, {Conroy}, {Conseil}, {Craig},
  {Cross}, {Cruz}, {D'Eugenio}, {Dencheva}, {Devillepoix}, {Dietrich},
  {Eigenbrot}, {Erben}, {Ferreira}, {Foreman-Mackey}, {Fox}, {Freij}, {Garg},
  {Geda}, {Glattly}, {Gondhalekar}, {Gordon}, {Grant}, {Greenfield}, {Groener},
  {Guest}, {Gurovich}, {Handberg}, {Hart}, {Hatfield-Dodds}, {Homeier},
  {Hosseinzadeh}, {Jenness}, {Jones}, {Joseph}, {Kalmbach}, {Karamehmetoglu},
  {Ka{l}uszy{'n}ski}, {Kelley}, {Kern}, {Kerzendorf}, {Koch}, {Kulumani},
  {Lee}, {Ly}, {Ma}, {MacBride}, {Maljaars}, {Muna}, {Murphy}, {Norman},
  {O'Steen}, {Oman}, {Pacifici}, {Pascual}, {Pascual-Granado}, {Patil},
  {Perren}, {Pickering}, {Rastogi}, {Roulston}, {Ryan}, {Rykoff}, {Sabater},
  {Sakurikar}, {Salgado}, {Sanghi}, {Saunders}, {Savchenko}, {Schwardt},
  {Seifert-Eckert}, {Shih}, {Jain}, {Shukla}, {Sick}, {Simpson},
  {Singanamalla}, {Singer}, {Singhal}, {Sinha}, {Sip{H{o}}cz}, {Spitler},
  {Stansby}, {Streicher}, {{{S}}umak}, {Swinbank}, {Taranu}, {Tewary},
  {Tremblay}, {Val-Borro}, {Van Kooten}, {Vasovi{'c}}, {Verma}, {de Miranda
  Cardoso}, {Williams}, {Wilson}, {Winkel}, {Wood-Vasey}, {Xue}, {Yoachim},
  {Zhang}, {Zonca}, \& {Astropy Project Contributors}}]{astropy:2022}
{Astropy Collaboration}, {Price-Whelan}, A.~M., {Lim}, P.~L., {et~al.} 2022,
  apj, 935, 167, \dodoi{10.3847/1538-4357/ac7c74}

\bibitem[{Backman {et~al.}(1992)Backman, Gillett, \&
  Witteborn}]{backman1992infrared}
Backman, D., Gillett, F., \& Witteborn, F. 1992, Astrophysical Journal, Part 1
  (ISSN 0004-637X), vol. 385, Feb. 1, 1992, p. 670-679., 385, 670

\bibitem[{Beichman {et~al.}(1988)Beichman, Neugebauer, Habing, Clegg, \&
  Chester}]{beichman1988infrared}
Beichman, C., Neugebauer, G., Habing, H., Clegg, P., \& Chester, T.~J. 1988,
  Infrared astronomical satellite (IRAS) catalogs and atlases. Volume 1:
  Explanatory supplement

\bibitem[{{Bessell} {et~al.}(1998){Bessell}, {Castelli}, \&
  {Plez}}]{1998A&A...333..231B}
{Bessell}, M.~S., {Castelli}, F., \& {Plez}, B. 1998, \aap, 333, 231

\bibitem[{Beust {et~al.}(1990)Beust, Lagrange-Henri, Vidal-Madjar, \&
  Ferlet}]{beust1990beta}
Beust, H., Lagrange-Henri, A., Vidal-Madjar, A., \& Ferlet, R. 1990, Astronomy
  and Astrophysics (ISSN 0004-6361), vol. 236, no. 1, Sept. 1990, p. 202-216.,
  236, 202

\bibitem[{Bonnefoy {et~al.}(2021)Bonnefoy, Milli, Menard, Delorme, Chomez,
  Bonavita, Lagrange, Vigan, Augereau, Beuzit, {et~al.}}]{bonnefoy2021narrow}
Bonnefoy, M., Milli, J., Menard, F., {et~al.} 2021, Astronomy \& Astrophysics,
  655, A62

\bibitem[{Booth {et~al.}(2013)Booth, Kennedy, Sibthorpe, Matthews, Wyatt,
  Duch{\^e}ne, Kavelaars, Rodriguez, Greaves, Koning,
  {et~al.}}]{booth2013resolved}
Booth, M., Kennedy, G., Sibthorpe, B., {et~al.} 2013, Monthly Notices of the
  Royal Astronomical Society, 428, 1263

\bibitem[{Brown {et~al.}(2021)Brown, Vallenari, Prusti, De~Bruijne, Babusiaux,
  Biermann, Creevey, Evans, Eyer, Hutton, {et~al.}}]{brown2021gaia}
Brown, A.~G., Vallenari, A., Prusti, T., {et~al.} 2021, Astronomy \&
  Astrophysics, 649, A1

\bibitem[{Burns {et~al.}(1979)Burns, Lamy, \& Soter}]{burns1979radiation}
Burns, J.~A., Lamy, P.~L., \& Soter, S. 1979, Icarus, 40, 1

\bibitem[{Castelli \& Kurucz(2004)}]{castelli2004new}
Castelli, F., \& Kurucz, R.~L. 2004, arXiv preprint astro-ph/0405087

\bibitem[{Cohen {et~al.}(2003)Cohen, Wheaton, \& Megeath}]{cohen2003spectral}
Cohen, M., Wheaton, W.~A., \& Megeath, S. 2003, The Astronomical Journal, 126,
  1090

\bibitem[{Cox(2015)}]{cox2015allen}
Cox, A.~N. 2015, Allen’s astrophysical quantities (Springer)

\bibitem[{Crotts {et~al.}(2024)Crotts, Matthews, Duch{\^e}ne, Esposito, Dong,
  Hom, Oppenheimer, Rice, Wolff, Chen, {et~al.}}]{crotts2024uniform}
Crotts, K.~A., Matthews, B.~C., Duch{\^e}ne, G., {et~al.} 2024, The
  Astrophysical Journal, 961, 245

\bibitem[{Cutri {et~al.}(2003)Cutri, Skrutskie, Van~Dyk, Beichman, Carpenter,
  Chester, Cambresy, Evans, Fowler, Gizis, {et~al.}}]{cutri20032mass}
Cutri, R., Skrutskie, M., Van~Dyk, S., {et~al.} 2003, The IRSA 2MASS All-Sky
  Point Source Catalog, \dodoi{10.26131/IRSA2}

\bibitem[{Cutri {et~al.}(2012)}]{cutri2012vizier}
Cutri, R., {et~al.} 2012, VizieR Online Data Catalog, II

\bibitem[{De~Rosa {et~al.}(2015)De~Rosa, Nielsen, Blunt, Graham, Konopacky,
  Marois, Pueyo, Rameau, Ryan, Wang, {et~al.}}]{de2015astrometric}
De~Rosa, R.~J., Nielsen, E.~L., Blunt, S.~C., {et~al.} 2015, The Astrophysical
  Journal Letters, 814, L3

\bibitem[{De~Rosa {et~al.}(2020)De~Rosa, Nguyen, Chilcote, Macintosh, Perrin,
  Konopacky, Wang, Duch{\^e}ne, Nielsen, Rameau, {et~al.}}]{de2020revised}
De~Rosa, R.~J., Nguyen, M.~M., Chilcote, J., {et~al.} 2020, Journal of
  Astronomical Telescopes, Instruments, and Systems, 6, 015006

\bibitem[{{De Rosa} {et~al.}(2020){De Rosa}, {Esposito}, {Gibbs}, {Bailey},
  {Fitzgerald}, {Chilcote}, {Duch{\^e}ne}, {Konopacky}, {Macintosh},
  {Millar-Blanchaer}, {Nguyen}, {Nielsen}, {Perrin}, {Rameau}, \&
  {Wang}}]{2020SPIE11447E..5AD}
{De Rosa}, R.~J., {Esposito}, T.~M., {Gibbs}, A., {et~al.} 2020, in Society of
  Photo-Optical Instrumentation Engineers (SPIE) Conference Series, Vol. 11447,
  Ground-based and Airborne Instrumentation for Astronomy VIII, ed. C.~J.
  {Evans}, J.~J. {Bryant}, \& K.~{Motohara}, 114475A,
  \dodoi{10.1117/12.2561071}

\bibitem[{Draine(2003)}]{draine2003interstellar}
Draine, B.~T. 2003, Annual Review of Astronomy and Astrophysics, 41, 241

\bibitem[{{Draine} \& {Lee}(1984)}]{Draine1984ApJ}
{Draine}, B.~T., \& {Lee}, H.~M. 1984, \apj, 285, 89, \dodoi{10.1086/162480}

\bibitem[{Engler {et~al.}(2020)Engler, Lazzoni, Gratton, Milli, Schmid,
  Chauvin, Kral, Pawellek, Th{\'e}bault, Boccaletti, {et~al.}}]{engler2020hd}
Engler, N., Lazzoni, C., Gratton, R., {et~al.} 2020, Astronomy \& Astrophysics,
  635, A19

\bibitem[{{Esposito} {et~al.}(2018){Esposito}, {Duch{\^e}ne}, {Kalas}, {Rice},
  {Choquet}, {Ren}, {Perrin}, {Chen}, {Arriaga}, {Chiang}, {Nielsen}, {Graham},
  {Wang}, {De Rosa}, {Follette}, {Ammons}, {Ansdell}, {Bailey}, {Barman},
  {Sebasti{\'a}n Bruzzone}, {Bulger}, {Chilcote}, {Cotten}, {Doyon},
  {Fitzgerald}, {Goodsell}, {Greenbaum}, {Hibon}, {Hung}, {Ingraham},
  {Konopacky}, {Larkin}, {Macintosh}, {Maire}, {Marchis}, {Marois}, {Mazoyer},
  {Metchev}, {Millar-Blanchaer}, {Oppenheimer}, {Palmer}, {Patience},
  {Poyneer}, {Pueyo}, {Rajan}, {Rameau}, {Rantakyr{\"o}}, {Ryan}, {Savransky},
  {Schneider}, {Sivaramakrishnan}, {Song}, {Soummer}, {Thomas}, {Wallace},
  {Ward-Duong}, {Wiktorowicz}, \& {Wolff}}]{esposito2018}
{Esposito}, T.~M., {Duch{\^e}ne}, G., {Kalas}, P., {et~al.} 2018, \aj, 156, 47,
  \dodoi{10.3847/1538-3881/aacbc9}

\bibitem[{Esposito {et~al.}(2020)Esposito, Kalas, Fitzgerald, Millar-Blanchaer,
  Duchene, Patience, Hom, Perrin, De~Rosa, Chiang,
  {et~al.}}]{esposito2020debris}
Esposito, T.~M., Kalas, P., Fitzgerald, M.~P., {et~al.} 2020, The Astronomical
  Journal, 160, 24

\bibitem[{Foreman-Mackey {et~al.}(2013)Foreman-Mackey, Hogg, Lang, \&
  Goodman}]{foreman2013emcee}
Foreman-Mackey, D., Hogg, D.~W., Lang, D., \& Goodman, J. 2013, Publications of
  the Astronomical Society of the Pacific, 125, 306

\bibitem[{Foreman-Mackey {et~al.}(2019)Foreman-Mackey, Farr, Sinha, Archibald,
  Hogg, Sanders, Zuntz, Williams, Nelson, de~Val-Borro,
  {et~al.}}]{foreman2019emcee}
Foreman-Mackey, D., Farr, W.~M., Sinha, M., {et~al.} 2019, arXiv preprint
  arXiv:1911.07688

\bibitem[{Fujiwara {et~al.}(2010)Fujiwara, Ishihara, Oyabu, Salama, Takita, \&
  Yamamura}]{fujiwaraakari}
Fujiwara, H., Ishihara, D., Oyabu, S., {et~al.} 2010, AKARI/IRC Catalogue

\bibitem[{Gaia {et~al.}(2018)Gaia, Brown, Vallenari, Prusti, De~Bruijne,
  Babusiaux, Juh{\'a}sz, Marschalk{\'o}, Marton, Moln{\'a}r,
  {et~al.}}]{gaia2018gaia}
Gaia, C., Brown, A., Vallenari, A., {et~al.} 2018, Astronomy \& Astrophysics,
  616

\bibitem[{{Gaia Collaboration}(2018)}]{2018yCat.1345....0G}
{Gaia Collaboration}. 2018, VizieR Online Data Catalog, I/345

\bibitem[{Goodman \& Weare(2010)}]{goodman2010ensemble}
Goodman, J., \& Weare, J. 2010, Communications in applied mathematics and
  computational science, 5, 65

\bibitem[{Grigorieva {et~al.}(2007)Grigorieva, Artymowicz, \&
  Th{\'e}bault}]{grigorieva2007collisional}
Grigorieva, A., Artymowicz, P., \& Th{\'e}bault, P. 2007, Astronomy \&
  Astrophysics, 461, 537

\bibitem[{Hales {et~al.}(2019)Hales, Gorti, Carpenter, Hughes, \&
  Flaherty}]{hales2019modeling}
Hales, A., Gorti, U., Carpenter, J.~M., Hughes, M., \& Flaherty, K. 2019, The
  Astrophysical Journal, 878, 113

\bibitem[{Henyey \& Greenstein(1941)}]{henyey1941diffuse}
Henyey, L.~G., \& Greenstein, J.~L. 1941, Astrophysical Journal, vol. 93, p.
  70-83 (1941)., 93, 70

\bibitem[{{H{\o}g} {et~al.}(2000){H{\o}g}, {Fabricius}, {Makarov}, {Urban},
  {Corbin}, {Wycoff}, {Bastian}, {Schwekendiek}, \&
  {Wicenec}}]{2000A&A...355L..27H}
{H{\o}g}, E., {Fabricius}, C., {Makarov}, V.~V., {et~al.} 2000, \aap, 355, L27

\bibitem[{Hog {et~al.}(2000)Hog, Fabricius, Makarov, Urban, Corbin, Wycoff,
  Bastian, Schwekendiek, \& Wicenec}]{hog2000tycho}
Hog, E., Fabricius, C., Makarov, V.~V., {et~al.} 2000, The Tycho-2 catalogue of
  the 2.5 million brightest stars, Tech. rep., Naval Observatory Washington DC

\bibitem[{Hom {et~al.}(2020)Hom, Patience, Esposito, Duch{\^e}ne, Worthen,
  Kalas, Jang-Condell, Saboi, Arriaga, Mazoyer, {et~al.}}]{hom2020first}
Hom, J., Patience, J., Esposito, T.~M., {et~al.} 2020, The Astronomical
  Journal, 159, 31

\bibitem[{Houk \& Fuentes-Williams(1982)}]{houk1982vol}
Houk, N., \& Fuentes-Williams, T. 1982, in Bulletin of the American
  Astronomical Society, Vol.~14, 615

\bibitem[{Hughes {et~al.}(2018)Hughes, Duch{\^e}ne, \&
  Matthews}]{hughes2018debris}
Hughes, A.~M., Duch{\^e}ne, G., \& Matthews, B.~C. 2018, Annual Review of
  Astronomy and Astrophysics, 56, 541

\bibitem[{Hung {et~al.}(2016)Hung, Bruzzone, Millar-Blanchaer, Wang, Arriaga,
  Metchev, Fitzgerald, Sivaramakrishnan, \& Perrin}]{hung2016gemini}
Hung, L.-W., Bruzzone, S., Millar-Blanchaer, M.~A., {et~al.} 2016, in
  Ground-based and Airborne Instrumentation for Astronomy VI, Vol. 9908, SPIE,
  1044--1056

\bibitem[{{Hunter}(2007)}]{matplotlib}
{Hunter}, J.~D. 2007, Computing in Science and Engineering, 9, 90,
  \dodoi{10.1109/MCSE.2007.55}

\bibitem[{Ishihara {et~al.}(2010)Ishihara, Onaka, Kataza, Salama, Alfageme,
  Cassatella, Cox, Garcia-Lario, Stephenson, Cohen,
  {et~al.}}]{ishihara2010AKARI}
Ishihara, D., Onaka, T., Kataza, H., {et~al.} 2010, Astronomy \& Astrophysics,
  514, A1

\bibitem[{Jones {et~al.}(2001)Jones, Oliphant, Peterson, {et~al.}}]{scipy}
Jones, E., Oliphant, T., Peterson, P., {et~al.} 2001, {SciPy}: Open source
  scientific tools for {Python}.
\newblock \url{http://www.scipy.org/}

\bibitem[{Kluyver {et~al.}(2016)Kluyver, Ragan-Kelley, P{\'e}rez, Granger,
  Bussonnier, Frederic, Kelley, Hamrick, Grout, Corlay, {et~al.}}]{jupyter}
Kluyver, T., Ragan-Kelley, B., P{\'e}rez, F., {et~al.} 2016, in ELPUB, 87--90

\bibitem[{Konopacky {et~al.}(2014)Konopacky, Thomas, Macintosh, Dillon,
  Sadakuni, Maire, Fitzgerald, Hinkley, Kalas, Esposito,
  {et~al.}}]{konopacky2014gemini}
Konopacky, Q.~M., Thomas, S.~J., Macintosh, B.~A., {et~al.} 2014, in
  Ground-based and Airborne Instrumentation for Astronomy V, Vol. 9147, SPIE,
  2585--2600

\bibitem[{Kral {et~al.}(2019)Kral, Marino, Wyatt, Kama, \&
  Matra}]{kral2019imaging}
Kral, Q., Marino, S., Wyatt, M.~C., Kama, M., \& Matra, L. 2019, Monthly
  Notices of the Royal Astronomical Society, 489, 3670

\bibitem[{Krist {et~al.}(2005)Krist, Ardila, Golimowski, Clampin, Ford,
  Illingworth, Hartig, Bartko, Ben{\'\i}tez, Blakeslee,
  {et~al.}}]{krist2005hubble}
Krist, J.~E., Ardila, D., Golimowski, D., {et~al.} 2005, The Astronomical
  Journal, 129, 1008

\bibitem[{{Lagrange} {et~al.}(2010){Lagrange}, {Bonnefoy}, {Chauvin}, {Apai},
  {Ehrenreich}, {Boccaletti}, {Gratadour}, {Rouan}, {Mouillet}, {Lacour}, \&
  {Kasper}}]{2010Sci...329...57L}
{Lagrange}, A.~M., {Bonnefoy}, M., {Chauvin}, G., {et~al.} 2010, Science, 329,
  57, \dodoi{10.1126/science.1187187}

\bibitem[{Lewis {et~al.}(2021)Lewis, Teng, \&
  rlopezucla}]{briley_l_lewis_2021_5029487}
Lewis, B.~L., Teng, E., \& rlopezucla. 2021, briley-lewis/elliptical-mask:
  Version 0.2, v0.2-beta,  Zenodo, \dodoi{10.5281/zenodo.5029487}

\bibitem[{Lieman-Sifry {et~al.}(2016)Lieman-Sifry, Hughes, Carpenter, Gorti,
  Hales, \& Flaherty}]{lieman2016debris}
Lieman-Sifry, J., Hughes, A.~M., Carpenter, J.~M., {et~al.} 2016, The
  Astrophysical Journal, 828, 25

\bibitem[{Macintosh {et~al.}(2014)Macintosh, Graham, Ingraham, Konopacky,
  Marois, Perrin, Poyneer, Bauman, Barman, Burrows, Cardwell, Chilcote,
  De~Rosa, Dillon, Doyon, Dunn, Erikson, Fitzgerald, Gavel, Goodsell, Hartung,
  Hibon, Kalas, Larkin, Maire, Marchis, Marley, McBride, Millar-Blanchaer,
  Morzinski, Norton, Oppenheimer, Palmer, Patience, Pueyo, Rantakyro, Sadakuni,
  Saddlemyer, Savransky, Serio, Soummer, Sivaramakrishnan, Song, Thomas,
  Wallace, Wiktorowicz, \& Wolff}]{Macintosh12661}
Macintosh, B., Graham, J.~R., Ingraham, P., {et~al.} 2014, Proceedings of the
  National Academy of Sciences, 111, 12661, \dodoi{10.1073/pnas.1304215111}

\bibitem[{Macintosh {et~al.}(2008)Macintosh, Graham, Palmer, Doyon, Dunn,
  Gavel, Larkin, Oppenheimer, Saddlemyer, Sivaramakrishnan,
  {et~al.}}]{macintosh2008gemini}
Macintosh, B.~A., Graham, J.~R., Palmer, D.~W., {et~al.} 2008, in Adaptive
  Optics Systems, Vol. 7015, International Society for Optics and Photonics,
  701518

\bibitem[{Mainzer {et~al.}(2011)Mainzer, Bauer, Grav, Masiero, Cutri, Dailey,
  Eisenhardt, McMillan, Wright, Walker, {et~al.}}]{mainzer2011preliminary}
Mainzer, A., Bauer, J., Grav, T., {et~al.} 2011, The Astrophysical Journal,
  731, 53

\bibitem[{Mann {et~al.}(2006)Mann, K{\"o}hler, Kimura, Cechowski, \&
  Minato}]{mann2006dust}
Mann, I., K{\"o}hler, M., Kimura, H., Cechowski, A., \& Minato, T. 2006, The
  Astronomy and Astrophysics Review, 13, 159

\bibitem[{Marino {et~al.}(2020)Marino, Flock, Henning, Kral, Matr{\`a}, \&
  Wyatt}]{marino2020population}
Marino, S., Flock, M., Henning, T., {et~al.} 2020, Monthly Notices of the Royal
  Astronomical Society, 492, 4409

\bibitem[{Marois {et~al.}(2006)Marois, Lafreniere, Doyon, Macintosh, \&
  Nadeau}]{marois2006angular}
Marois, C., Lafreniere, D., Doyon, R., Macintosh, B., \& Nadeau, D. 2006, The
  Astrophysical Journal, 641, 556

\bibitem[{Mellon {et~al.}(2019)Mellon, Mamajek, Zwintz, David, Stuik, Talens,
  Dorval, Burggraaff, Kenworthy, Bailey~III, {et~al.}}]{mellon2019discovery}
Mellon, S.~N., Mamajek, E.~E., Zwintz, K., {et~al.} 2019, The Astrophysical
  Journal, 870, 36

\bibitem[{Millar-Blanchaer {et~al.}(2015)Millar-Blanchaer, Graham, Pueyo,
  Kalas, Dawson, Wang, Perrin, Macintosh, Ammons, Barman,
  {et~al.}}]{millar2015beta}
Millar-Blanchaer, M.~A., Graham, J.~R., Pueyo, L., {et~al.} 2015, The
  Astrophysical Journal, 811, 18

\bibitem[{Milli {et~al.}(2012)Milli, Mouillet, Lagrange, Boccaletti, Mawet,
  Chauvin, \& Bonnefoy}]{milli2012impact}
Milli, J., Mouillet, D., Lagrange, A.-M., {et~al.} 2012, Astronomy \&
  Astrophysics, 545, A111

\bibitem[{Mo{\'o}r {et~al.}(2017)Mo{\'o}r, Cur{\'e}, K{\'o}sp{\'a}l,
  {\'A}brah{\'a}m, Csengeri, Eiroa, Gunawan, Henning, Hughes, Juh{\'a}sz,
  {et~al.}}]{moor2017molecular}
Mo{\'o}r, A., Cur{\'e}, M., K{\'o}sp{\'a}l, {\'A}., {et~al.} 2017, The
  Astrophysical Journal, 849, 123

\bibitem[{Mo{\'o}r {et~al.}(2019)Mo{\'o}r, Kral, {\'A}brah{\'a}m,
  K{\'o}sp{\'a}l, Dutrey, Di~Folco, Hughes, Juh{\'a}sz, Pascucci, \&
  Pawellek}]{moor2019new}
Mo{\'o}r, A., Kral, Q., {\'A}brah{\'a}m, P., {et~al.} 2019, The Astrophysical
  Journal, 884, 108

\bibitem[{Morales {et~al.}(2011)Morales, Rieke, Werner, Bryden, Stapelfeldt, \&
  Su}]{morales2011common}
Morales, F.~Y., Rieke, G., Werner, M., {et~al.} 2011, The Astrophysical Journal
  Letters, 730, L29

\bibitem[{Murakami {et~al.}(2007)Murakami, Baba, Barthel, Clements, Cohen, Doi,
  Enya, Figueredo, Fujishiro, Fujiwara, {et~al.}}]{murakami2007infrared}
Murakami, H., Baba, H., Barthel, P., {et~al.} 2007, Publications of the
  Astronomical Society of Japan, 59, S369

\bibitem[{Neugebauer {et~al.}(2019)Neugebauer, Habing, van Duinen,
  {et~al.}}]{neugebauer2019IRAS}
Neugebauer, G., Habing, H., van Duinen, R., {et~al.} 2019, IPAC, doi, 10,
  \dodoi{10.26131/IRSA4}

\bibitem[{{Neugebauer} {et~al.}(1984){Neugebauer}, {Habing}, {van Duinen},
  {Aumann}, {Baud}, {Beichman}, {Beintema}, {Boggess}, {Clegg}, {de Jong},
  {Emerson}, {Gautier}, {Gillett}, {Harris}, {Hauser}, {Houck}, {Jennings},
  {Low}, {Marsden}, {Miley}, {Olnon}, {Pottasch}, {Raimond}, {Rowan-Robinson},
  {Soifer}, {Walker}, {Wesselius}, \& {Young}}]{1984ApJ...278L...1N}
{Neugebauer}, G., {Habing}, H.~J., {van Duinen}, R., {et~al.} 1984, \apjl, 278,
  L1, \dodoi{10.1086/184209}

\bibitem[{Nielsen {et~al.}(2019)Nielsen, De~Rosa, Macintosh, Wang, Ruffio,
  Chiang, Marley, Saumon, Savransky, Ammons, {et~al.}}]{nielsen2019gemini}
Nielsen, E.~L., De~Rosa, R.~J., Macintosh, B., {et~al.} 2019, The Astronomical
  Journal, 158, 13

\bibitem[{Pawellek \& Krivov(2015)}]{pawellek2015dust}
Pawellek, N., \& Krivov, A.~V. 2015, Monthly Notices of the Royal Astronomical
  Society, 454, 3207

\bibitem[{Pawellek {et~al.}(2021)Pawellek, Wyatt, Matra, Kennedy, \&
  Yelverton}]{pawellek202175}
Pawellek, N., Wyatt, M., Matra, L., Kennedy, G., \& Yelverton, B. 2021, Monthly
  Notices of the Royal Astronomical Society, 502, 5390

\bibitem[{Pearce {et~al.}(2024)Pearce, Krivov, Sefilian, Jankovic, L{\"o}hne,
  Morgner, Wyatt, Booth, \& Marino}]{pearce2024effect}
Pearce, T.~D., Krivov, A.~V., Sefilian, A.~A., {et~al.} 2024, Monthly Notices
  of the Royal Astronomical Society, 527, 3876

\bibitem[{Pecaut \& Mamajek(2016)}]{pecaut2016star}
Pecaut, M.~J., \& Mamajek, E.~E. 2016, Monthly Notices of the Royal
  Astronomical Society, 461, 794

\bibitem[{{Perez} \& {Granger}(2007)}]{ipython}
{Perez}, F., \& {Granger}, B.~E. 2007, Computing in Science and Engineering, 9,
  21, \dodoi{10.1109/MCSE.2007.53}

\bibitem[{P{\'e}ricaud {et~al.}(2017)P{\'e}ricaud, Di~Folco, Dutrey,
  Guilloteau, \& Pi{\'e}tu}]{pericaud2017hybrid}
P{\'e}ricaud, J., Di~Folco, E., Dutrey, A., Guilloteau, S., \& Pi{\'e}tu, V.
  2017, Astronomy \& Astrophysics, 600, A62

\bibitem[{Perrin {et~al.}(2010)Perrin, Graham, Larkin, Wiktorowicz, Maire,
  Thibault, Fitzgerald, Doyon, Macintosh, Gavel, {et~al.}}]{perrin2010imaging}
Perrin, M.~D., Graham, J.~R., Larkin, J.~E., {et~al.} 2010, in Adaptive Optics
  Systems II, Vol. 7736, SPIE, 1964--1972

\bibitem[{Perrin {et~al.}(2014)Perrin, Maire, Ingraham, Savransky,
  Millar-Blanchaer, Wolff, Ruffio, Wang, Draper, Sadakuni,
  {et~al.}}]{perrin2014gemini}
Perrin, M.~D., Maire, J., Ingraham, P., {et~al.} 2014, Ground-based and
  Airborne Instrumentation for Astronomy V, 9147, 1168

\bibitem[{Perrin {et~al.}(2015)Perrin, Duchene, Millar-Blanchaer, Fitzgerald,
  Graham, Wiktorowicz, Kalas, Macintosh, Bauman, Cardwell,
  {et~al.}}]{perrin2015polarimetry}
Perrin, M.~D., Duchene, G., Millar-Blanchaer, M., {et~al.} 2015, The
  Astrophysical Journal, 799, 182

\bibitem[{Perrin {et~al.}(2016)Perrin, Ingraham, Follette, Maire, Wang,
  Savransky, Arriaga, Bailey, Bruzzone, Chilcote, {et~al.}}]{perrin2016gemini}
Perrin, M.~D., Ingraham, P., Follette, K.~B., {et~al.} 2016, in Ground-based
  and Airborne Instrumentation for Astronomy VI, Vol. 9908, SPIE, 1010--1022

\bibitem[{{Pickles} \& {Depagne}(2010)}]{2010PASP..122.1437P}
{Pickles}, A., \& {Depagne}, {\'E}. 2010, \pasp, 122, 1437,
  \dodoi{10.1086/657947}

\bibitem[{Pinte(2022)}]{pymcfost}
Pinte, C. 2022, pymcfost, \url{https://github.com/cpinte/pymcfost},  GitHub

\bibitem[{Pinte {et~al.}(2009)Pinte, Harries, Min, Watson, Dullemond, Woitke,
  M{\'e}nard, \& Dur{\'a}n-Rojas}]{pinte2009benchmark}
Pinte, C., Harries, T., Min, M., {et~al.} 2009, Astronomy \& Astrophysics, 498,
  967

\bibitem[{Pinte {et~al.}(2006)Pinte, M{\'e}nard, Duch{\^e}ne, \&
  Bastien}]{pinte2006monte}
Pinte, C., M{\'e}nard, F., Duch{\^e}ne, G., \& Bastien, P. 2006, Astronomy \&
  Astrophysics, 459, 797

\bibitem[{Ratzenb{\"o}ck {et~al.}(2023)Ratzenb{\"o}ck, Gro{\ss}schedl, Alves,
  Miret-Roig, Bomze, Forbes, Goodman, Hacar, Lin, Meingast,
  {et~al.}}]{ratzenbock2023star}
Ratzenb{\"o}ck, S., Gro{\ss}schedl, J.~E., Alves, J., {et~al.} 2023, Astronomy
  \& Astrophysics, 678, A71

\bibitem[{Ren {et~al.}(2018)Ren, Pueyo, Zhu, Debes, \&
  Duch{\^e}ne}]{ren2018non}
Ren, B., Pueyo, L., Zhu, G.~B., Debes, J., \& Duch{\^e}ne, G. 2018, The
  Astrophysical Journal, 852, 104

\bibitem[{Rizzuto {et~al.}(2012)Rizzuto, Ireland, \& Zucker}]{rizzuto2012WISE}
Rizzuto, A.~C., Ireland, M.~J., \& Zucker, D.~B. 2012, Monthly Notices of the
  Royal Astronomical Society: Letters, 421, L97

\bibitem[{Rodrigo \& Solano(2020)}]{rodrigo2020svo}
Rodrigo, C., \& Solano, E. 2020, in XIV. 0 Scientific Meeting (virtual) of the
  Spanish Astronomical Society, 182

\bibitem[{Roques {et~al.}(1994)Roques, Scholl, Sicardy, \&
  Smith}]{roques1994there}
Roques, F., Scholl, H., Sicardy, B., \& Smith, B.~A. 1994, Icarus, 108, 37

\bibitem[{Schmid {et~al.}(2006)Schmid, Joos, \& Tschan}]{schmid2006limb}
Schmid, H., Joos, F., \& Tschan, D. 2006, Astronomy \& Astrophysics, 452, 657

\bibitem[{Sivaramakrishnan \&
  Oppenheimer(2006)}]{sivaramakrishnan2006astrometry}
Sivaramakrishnan, A., \& Oppenheimer, B.~R. 2006, The Astrophysical Journal,
  647, 620

\bibitem[{Skrutskie {et~al.}(2006)Skrutskie, Cutri, Stiening, Weinberg,
  Schneider, Carpenter, Beichman, Capps, Chester, Elias,
  {et~al.}}]{skrutskie2006two}
Skrutskie, M., Cutri, R., Stiening, R., {et~al.} 2006, The Astronomical
  Journal, 131, 1163

\bibitem[{{Skrutskie} {et~al.}(2006){Skrutskie}, {Cutri}, {Stiening},
  {Weinberg}, {Schneider}, {Carpenter}, {Beichman}, {Capps}, {Chester},
  {Elias}, {Huchra}, {Liebert}, {Lonsdale}, {Monet}, {Price}, {Seitzer},
  {Jarrett}, {Kirkpatrick}, {Gizis}, {Howard}, {Evans}, {Fowler}, {Fullmer},
  {Hurt}, {Light}, {Kopan}, {Marsh}, {McCallon}, {Tam}, {Van Dyk}, \&
  {Wheelock}}]{2006AJ....131.1163S}
{Skrutskie}, M.~F., {Cutri}, R.~M., {Stiening}, R., {et~al.} 2006, \aj, 131,
  1163, \dodoi{10.1086/498708}

\bibitem[{Soummer {et~al.}(2012)Soummer, Pueyo, \& Larkin}]{soummer2012KLIP}
Soummer, R., Pueyo, L., \& Larkin, J. 2012, The Astrophysical Journal Letters,
  755, L28

\bibitem[{Stark {et~al.}(2014)Stark, Schneider, Weinberger, Debes, Grady,
  Jang-Condell, \& Kuchner}]{stark2014revealing}
Stark, C.~C., Schneider, G., Weinberger, A.~J., {et~al.} 2014, The
  Astrophysical Journal, 789, 58

\bibitem[{{Stolker} {et~al.}(2016){Stolker}, {Dominik}, {Min}, {Garufi},
  {Mulders}, \& {Avenhaus}}]{2016A&A...596A..70S}
{Stolker}, T., {Dominik}, C., {Min}, M., {et~al.} 2016, \aap, 596, A70,
  \dodoi{10.1051/0004-6361/201629098}

\bibitem[{Tanab{\'E} {et~al.}(2008)Tanab{\'E}, Sakon, Cohen, Wada, Ita, Ohyama,
  Oyabu, Uemizu, Takagi, Ishihara, {et~al.}}]{tanabe2008absolute}
Tanab{\'E}, T., Sakon, I., Cohen, M., {et~al.} 2008, Publications of the
  Astronomical Society of Japan, 60, S375

\bibitem[{Th{\'e}bault(2009)}]{thebault2009vertical}
Th{\'e}bault, P. 2009, Astronomy \& Astrophysics, 505, 1269

\bibitem[{Tokunaga \& Vacca(2005)}]{tokunaga2005mauna}
Tokunaga, A., \& Vacca, W. 2005, Publications of the Astronomical Society of
  the Pacific, 117, 421

\bibitem[{{van der Walt} {et~al.}(2011){van der Walt}, {Colbert}, \&
  {Varoquaux}}]{numpy}
{van der Walt}, S., {Colbert}, S.~C., \& {Varoquaux}, G. 2011, Computing in
  Science and Engineering, 13, 22, \dodoi{10.1109/MCSE.2011.37}

\bibitem[{Wang {et~al.}(2015)Wang, Ruffio, De~Rosa, Aguilar, Wolff, \&
  Pueyo}]{wang2015pyklip}
Wang, J.~J., Ruffio, J.-B., De~Rosa, R.~J., {et~al.} 2015, Astrophysics Source
  Code Library, ascl

\bibitem[{Wang {et~al.}(2014)Wang, Rajan, Graham, Savransky, Ingraham,
  Ward-Duong, Patience, De~Rosa, Bulger, Sivaramakrishnan,
  {et~al.}}]{wang2014gemini}
Wang, J.~J., Rajan, A., Graham, J.~R., {et~al.} 2014, in Ground-based and
  Airborne Instrumentation for Astronomy V, Vol. 9147, SPIE, 1678--1692

\bibitem[{Wang {et~al.}(2018)Wang, Perrin, Savransky, Arriaga, Chilcote,
  De~Rosa, Millar-Blanchaer, Marois, Rameau, Wolff,
  {et~al.}}]{wang2018automated}
Wang, J.~J., Perrin, M.~D., Savransky, D., {et~al.} 2018, Journal of
  Astronomical Telescopes, Instruments, and Systems, 4, 018002

\bibitem[{Wiktorowicz {et~al.}(2014)Wiktorowicz, Millar-Blanchaer, Perrin,
  Graham, Fitzgerald, Maire, Ingraham, Savransky, Macintosh, Thomas,
  {et~al.}}]{wiktorowicz2014gemini}
Wiktorowicz, S.~J., Millar-Blanchaer, M., Perrin, M.~D., {et~al.} 2014, in
  Ground-based and Airborne Instrumentation for Astronomy V, Vol. 9147, SPIE,
  2574--2584

\bibitem[{Wolff {et~al.}(2017{\natexlab{a}})Wolff, Perrin, Ren, \&
  Pinte}]{mcpy}
Wolff, S., Perrin, M., Ren, B., \& Pinte, C. 2017{\natexlab{a}},
  swolff9/mcfost-python: mcfost-python-v1.0,  Zenodo,
  \dodoi{10.5281/ZENODO.839863}

\bibitem[{Wolff {et~al.}(2017{\natexlab{b}})Wolff, Perrin, Stapelfeldt,
  Duch{\^e}ne, M{\'e}nard, Padgett, Pinte, Pueyo, \& Fischer}]{wolff2017hubble}
Wolff, S.~G., Perrin, M.~D., Stapelfeldt, K., {et~al.} 2017{\natexlab{b}}, The
  Astrophysical Journal, 851, 56

\bibitem[{Wright {et~al.}(2019)Wright, Eisenhardt, Mainzer, Ressler, \&
  Cutri}]{wright2019allWISE}
Wright, E.~L., Eisenhardt, P.~R., Mainzer, A.~K., Ressler, M.~E., \& Cutri,
  R.~M. 2019, IPAC, doi, 10, \dodoi{10.26131/IRSA1}

\bibitem[{Wright {et~al.}(2010)Wright, Eisenhardt, Mainzer, Ressler, Cutri,
  Jarrett, Kirkpatrick, Padgett, McMillan, Skrutskie,
  {et~al.}}]{wright2010wide}
Wright, E.~L., Eisenhardt, P.~R., Mainzer, A.~K., {et~al.} 2010, The
  Astronomical Journal, 140, 1868

\bibitem[{Wyatt(2008)}]{wyatt2008evolution}
Wyatt, M.~C. 2008, Annu. Rev. Astron. Astrophys., 46, 339

\end{thebibliography}

\end{document}